\UseRawInputEncoding
\documentclass[twocolumn]{aastex62}
\usepackage{natbib,amsmath,fleqn,graphicx,amssymb}
\usepackage{times,bm}
\usepackage{url}
\usepackage{txfonts}
\usepackage{epsfig,mathptmx}

\def\mnras{MNRAS}
\def\apj{ApJ}
\def\aj{AJ}
\def\aap{A\&A}
\def\cjaa{Chin. J. Astron. Astrophys.}

\def\kms{\mathrm{km\,s}^{-1}}
\def\kpc{\mathrm{kpc}}
\def\kmskpc{\mathrm{km\,s^{-1}\,kpc^{-1}}}
\def\masyr{\mathrm{mas\,yr}^{-1}}
\def\feh{\rm{[Fe/H]}}
\def\teff{T_{\rm{eff}}}
\def\logg{\rm{log}\emph{g}}
\def\vlos{v_{\rm{los}}}
\def\vbreath{V_{\rm{breath}}}
\def\vbend{V_{\rm{bend}}}
\def\tilt{\alpha_{\rm{tilt}}}

\shorttitle{Vertical structure of Galactic disk kinematics}
\shortauthors{Ding et al.}

\begin{document}

\title{Vertical structure of Galactic disk kinematics from LAMOST K giants}

\author{Ping-Jie Ding}
\affiliation{CAS Key Laboratory of Optical Astronomy, National Astronomical Observatories, Beijing 100101, China}

\author[0000-0002-0642-5689]{Xiang-Xiang Xue}
\affiliation{CAS Key Laboratory of Optical Astronomy, National Astronomical Observatories, Beijing 100101, China}

\author[0000-0003-1972-0086]{Chengqun Yang}
\affiliation{CAS Key Laboratory of Optical Astronomy, National Astronomical Observatories, Beijing 100101, China}
\affiliation{School of Astronomy and Space Science, University of Chinese Academy of Sciences, 19A Yuquan Road, Shijingshan District, Beijing 100049, P.R.China}
\affiliation{Shanghai Astronomical Observatory, 80 Nandan Road, Shanghai 200030, People¡¯s Republic of China}

\author[0000-0002-8980-945X]{Gang Zhao}
\affiliation{CAS Key Laboratory of Optical Astronomy, National Astronomical Observatories, Beijing 100101, China}
\affiliation{School of Astronomy and Space Science, University of Chinese Academy of Sciences, 19A Yuquan Road, Shijingshan District, Beijing 100049, P.R.China}

\author{Lan Zhang}
\affiliation{CAS Key Laboratory of Optical Astronomy, National Astronomical Observatories, Beijing 100101, China}

\author{Zi Zhu}
\affiliation{School of Astronomy and Space Science, Nanjing University, Nanjing 210023, China}
\affiliation{Key Laboratory for Modern Astronomy and Astrophysics (Nanjing University), Ministry of Education, Nanjing 210023, China}

\begin{abstract}
We examine the vertical structure of Galactic disk kinematics over a Galactocentric radial distance range of $R=5$-15 kpc and up to 3 kpc away from the Galactic plane, using the K-type giants surveyed by LAMOST. Based on robust measurements of three-dimensional velocity moments, a wobbly disk is detected in a phenomenological sense. An outflow dominates the radial motion of the inner disk, while in the outer disk there exist alternate outward and inward flows. The vertical bulk velocities is a combination of breathing and bending modes. A contraction-like breathing mode with amplitudes increasing with the distance to the plane and an upward bending mode dominate the vertical motion outside $R_0$, and there are reversed breathing mode and bending mode at $R<R_0$, with amplitudes much smaller than those outside $R_0$. The mean azimuthal velocity decreases with the increasing distance to the plane, with gradients shallower for larger $R$. Stars in the south disk are rotating faster than stars in the north. The velocity ellipsoid orientation differs between different $R$: in the range of $5<R<9\,\kpc$, the gradient of the tilt angle with respect to $\arctan(Z/R)$ decreases from $\sim0.83$ for the inner disk to $\sim0.52$ for the outer disk; within $9<R<15\,\kpc$, the tilt of velocity ellipsoid deviates from vertical antisymmetry. A clear flaring signature is found for both north and south disks based on the observed vertical structures of velocity ellipsoid.
\end{abstract}


\section{Introduction}

Our host galaxy, the Milky Way, is a typical disk galaxy. The vertical structure of the disk kinematics is important for our understanding of the Galactic formation and evolution. In the classical characterization of the Galactic disk, the distribution of stellar kinematics along the vertical distance to the Galactic plane is monotonous and symmetric. Owing to observations carried out by large surveys in recent years such as the Sloan Digital Sky Survey (SDSS; \citealt{York00}), the RAdial Velocity Experiment (RAVE; \citealt{Steinmetz06}), the Large Aperture Multi-Object Fibre Spectroscopic Telescope (LAMOST; \citealt{Cui12}), and the {\it Gaia} mission (\citealt{Gaia Collaboration16}), more and more details in the velocity profiles have become apparent.

The bulk motions in the Galactocentric radial and vertical directions and their variations with respect to the vertical height above and below the plane have been studied using different tracers. Within the disk, there is evidence of a wobbly radial velocity ($V_R$) along the distance to the plane. Investigating velocity profiles for F-type stars sampled from the LAMOST data (\citealt{Zhao06}; \citealt{Cui12}; \citealt{Zhao12}) within Galactocentric radii $7.8<R<9.8\,\kpc$ and $\pm 2\,\kpc$ from the plane, \citet{Carlin13} detected a negative mean radial velocity near the plane, corresponding to an inward radial flow. They also found that the mean of $V_R$ increases with the increasing of $|Z|$. Based on RAVE (\citealt{Steinmetz06}; \citealt{Siebert11}) red-clump stars at $|Z|<2\,\kpc$ and $6<R<10\,\kpc$, \citet{Williams13} found an outflow with $V_R=8-10\,\kms$ for $0<Z<1\,\kpc$ and an inward motion with $V_R=-10\,\kms$ at $R=9\,\kpc$ and $-1<Z<-0.5\,\kpc$. \citet{Wang18} presented an analysis of kinematics of K giant stars selected from the LAMOST catalog within $|Z|<2\,\kpc$ and found that the $V_R$ above the plane is higher than that below the plane at $R\sim10-11\,\kpc$. Similar north-south asymmetries in $V_R$ in the outer disk were also detected by \citet{Wang19} and \citet{Wang20} based on the LAMOST red clump stars. Sampling {\it Gaia}-LAMOST AFGK dwarf stars restricted to 1 kpc from the Sun, \citet{Ding19} found a $V_R$ profile with a gradient $dV_R/dZ\sim29\,\kmskpc$ across the plane.

The vertical motion of the disk can be decomposed into a breathing mode motion and a bending mode motion. The former is defined by a vertical pattern with odd parity in the $V_Z$ distribution with respect to $Z$, and the latter is a pattern with even parity in the $Z$-$V_Z$ plane \citep{Weinberg91}. Both the breathing and the bending modes have been detected in the solar neighborhood, though with differences between works. \citet{Widrow12} found that the vertical motion of the SDSS-SEGUE(\citealt{Aihara11}; \citealt{Yanny09}) stars resemble that of a breathing mode motion, in which stars above the plane are on average moving upwards, while those below the plane, downwards. \citet{Carlin13} argued that stars both above and below the plane are moving towards the plane, as a compression movement. A more complex structure was detected by \citet{Williams13}. They found opposite breathing modes inside and outside the solar circle: the $V_Z$ field implies a rarefaction at $R<R_0$, whereas a contraction at $R>R_0$. The inner-disk breathing mode found by \citet{Wang18} is the inverse of that found by \citet{Williams13}. Moreover, they detected a bending motion in the range of $6<R<13\,\kpc$, which is downward interior to the solar circle and upward at $R\gtrsim9\,\kpc$. The findings of \citet{Wang19} and \citet{Wang20} agree with a significant upward bending motion in the outer disk. Using data from RAVE DR5 \citep{Kunder17} and the Tycho-Gaia astrometric solution catalogue \citep{Michalik15}, \citet{Carrillo18} claimed that a rarefaction-like breathing mode dominates the $V_Z$ pattern inside $R_0$, while the pattern of the outer disk is dominated by a downward bending mode within $|Z|\lesssim1\,\kpc$. A similar combination of vertical modes was reported by \citet{Carrillo19} based on {\it Gaia} DR2 astrometric and line-of-sight velocity information, while a positive bending mode was observed at $R\thickapprox13\,\kpc$. In the work of \citet{Lopez-Corredoira20}, the authors found that both the inner and the outer disks present a combination of breathing and bending mode motions with {\it Gaia} DR2: a strong bending mode exists across the disk, the velocity of which changes from negative at $R<R_0$ to positive at $R>R_0$ similar to the pattern found by \citet{Wang18}; a relatively weak breathing mode shows a smooth distribution over the disk, with amplitudes highly dependent on the distance to the plane. Recently, \citet{Gaia Collaboration21} adopted proper motions of stars in the direction of the Galactic anticenter from the {\it Gaia} EDR3 (\citealt{Gaia Collaboration20}; \citealt{Lindegren20b}) and detected upward vertical motions in both north and south disks. They found that the $Z>0$ stars exhibit faster vertical motions than the $Z<0$ stars by up to $\sim2\,\kms$ at $R<10\,\kpc$, while at $R>10\,\kpc$ the vertical velocities of $Z<0$ stars are larger than those of $Z>0$ stars by up to $\sim6\,\kms$, implying a combination of an alternate breathing mode motion and an upward bending mode motion.

Compared to the behaviors of $V_R$ and $V_Z$, the vertical structure of the azimuthal velocities ($V_\phi$) is more regular, with a negative $V_\phi$-$|Z|$ trend on both sides of the disk (\citealt{Smith12}; \citealt{Karaali14}; \citealt{Recio-Blanco14}; \citealt{Guiglion15}; \citealt{Jing16}; \citealt{Carrillo18}). The vertical gradient in the azimuthal velocities is dependent on the tracers. For instance, sampling solar neighborhood dwarf stars in the range of $7<R<9\,\kpc$ and $|Z|<2\,\kpc$, \citet{Smith12} found that the gradient of the mean $V_\phi$ with respect to $|Z|$ is around $-15$ to $-40\,\kmskpc$. Using 130043 F/G-type dwarf stars within $6.5<R<9.5\,\kpc$ and $0.1<|Z|<3\,\kpc$ from the LAMOST survey and the SDSS, \citet{Jing16} obtained the gradient as $-18.5\,\kmskpc$ for the metallicity range $\feh>-0.1$ and $-14.2\,\kmskpc$ for $-0.8<\feh<-0.6$. In most of the previous research, the patterns of $V_\phi$ were approximately symmetric about the Galactic plane. However, some small-amplitude asymmetric structures have been detected in recent studies. Based on proper motions of the red clump giants from the PPMXL survey \citep{Roeser10}, \citet{Lopez-Corredoira14} found that the rotation speed of the south disk is higher than that in the north in the range of $8<R<13\,\kpc$, though with large error bars. \citet{Wang18} claimed that the K giants within $R\geqslant10\,\kpc$ and $Z\geqslant0.5\,\kpc$ exhibit larger rotation speed in the south disk than in the north, while at $R\thickapprox8-9\,\kpc$ and $|Z|\leqslant0.5\,\kpc$ northern stars are rotating faster than the southern. In the work of \citet{Ding19}, the authors found that dwarfs at $-0.5<Z<0\,\kpc$ are rotating slightly faster than those at $0<Z<0.5\,\kpc$. The faster-rotating of the southern stars was also evidenced by \citet{Gaia Collaboration21}. They claimed that the rotation of $Z<0$ stars leads that of $Z\geq0$ stars at $R\gtrsim11$ kpc, typically by up to $10\,\kms$.

In addition to the bulk motions, the velocity ellipsoid as a function of the vertical height provides information on the structure and evolution process of the disk. The velocity dispersion is known to be caused by the disk heating due to fluctuations in the gravitational field. An increment with the distance to the plane of all the velocity dispersions (i.e. $\sigma_R$, $\sigma_Z$ and $\sigma_\phi$) has been recognized in earlier works (e.g. \citealt{Soubiran03}; \citealt{MoniBidin12}; \citealt{Smith12}; \citealt{Karaali14}; \citealt{Ding19}; \citealt{Mackereth19}), indicating that the heating process is stronger for higher disk stars. Moreover, it has been found that the covariance between $V_R$ and $V_Z$, often parameterized by the tilt angle ($\tilt$) of the orientation of the velocity ellipsoid, is antisymmetric about the plane, and the orientation of the velocity ellipsoid stays between an horizontal one ($\tilt\equiv0$) and a spherical one ($\tilt=\arctan(Z/R)$). In the work of \citet{Binney14}, the authors found $\tilt\sim0.8\arctan(Z/R)$ by using RAVE stars lying within $\sim2\,\kpc$ of the Sun, suggesting a near-spherical orientation of the velocity ellipsoid. The findings of \citet{Budenbender15} and \citet{Ding19} based on nearby dwarf stars, which gave $\tilt=(0.90\pm0.04)\arctan(Z/R_0)-(0.01\pm0.005)$ and $\tilt=(1.11\pm0.11)\arctan(Z/R_0)-(0.0069\pm0.0034)$ respectively, agree with a spherical alignment of the tilt in the solar vicinity. Using data from the {\it Gaia} RVS catalog \citep{Cropper18} within 5 kpc of the Sun, \citet{Everall19} found the gradient of $\tilt$ with $\arctan(Z/R)$ is around 0.95. They also claimed that the tilt is closer to a spherical orientation at larger $R$ and higher $|Z|$. To the almost contrary, \citet{Hagen19} found a (near) spherical orientation of the velocity ellipsoid at $R\lesssim7\,\kpc$ and a more cylindrical tilt beyond $R\gtrsim9\,\kpc$ by sampling stars within $4\lesssim R\lesssim13\,\kpc$ and $|Z|\lesssim3.5\,\kpc$ from {\it Gaia} DR2.

The structural complexity in the vertical distribution of stellar velocities and the inconsistency in the earlier findings call for a reanalysis of vertical profiles of velocity moments in a phenomenological sense, by using favorable tracers obtained from developed observations. In this work, we focus on how the measured three-dimensional velocity moments modify the vertical patterns of disk kinematics, by sampling K giant stars with fundamental stellar parameters and line-of-sight velocities from the recent released LAMOST catalog. With the help of the high-precision measurements of proper motions provided by the {\it Gaia} EDR3, we examine in detail the variation of velocity moments with the distance to the plane in a large volume around the Sun and analyze the Galactic structure inferred from the observed kinematics.

In Section 2, we introduce our sample of K giant stars, for which we have LAMOST line-of-sight velocities, distances, and metallicities, matched to {\it Gaia} EDR3 proper motions. Section 3 gives the vertical distribution of velocity moments. The difference between the inner and outer disks as well as the dependence of stellar kinematics on the metallicity will also be presented. In Section 4, we provide a measurement of the flaring strength of the stellar disk by using the results in Section 3. We draw our discussion in Section 5 and present a summary of our results in Section 6.

\section{Sample selection}

In this work, we focus on the characterisation of the vertical structure of velocity moments across the Galactic disk. The K-type giants are a kind of long-lived objects with high luminosities. The large sample size and the wide spatial coverage of the observed K giants make them feasible tracers for the Galactic kinematics. We sample the K giants from the catalog of LAMOST DR8, which contains observations from October 24, 2011 to May 27, 2020. The K giants are selected based on the surface gravities ($\logg$) and effective temperatures ($\teff$) measured by the LAMOST team. The selection criteria follow the conditions given by \citet{Liu14} such that $4000\,\rm{K}<\teff<4600\,\rm{K}$ with $\logg<3.5$ and $4600\,\rm{K}<\teff<5600\,\rm{K}$ with $\logg<4$. We obtain the metallicities ($\feh$) and the line-of-sight velocities ($\vlos$) from the LAMOST data. We reject stars with $\feh\leqslant-1$ to exclude likely halo stars. It has been noted that there exists a systematic offset in the measured $\vlos$ of around $5-7\kms$ (\citealt{Gao15}; \citealt{Luo15}; \citealt{Tian15}; \citealt{Jing16}; \citealt{SA17}), which needs to be added to $\vlos$. We select stars from the LAMOST sample that also have measurements of line-of-sight velocities in the {\it Gaia} data to determine the offset. The {\it Gaia} EDR3 contains line-of-sight velocities for more than 7 million sources (\citealt{Cropper18}, \citealt{Gaia Collaboration20}), with systematic bias at a level of $0.1\,\kms$ \citep{Katz19}. We use the {\it Gaia} line-of-sight velocities satisfying $\rm{RV_{-}NB_{-}TRANSITS}>5$ as calibrators, and find that the offset in the LAMOST measurement is $5.34\,\kms$ (see Fig. 1).

\begin{figure}
\centering
\centerline{
\includegraphics[width=0.7\hsize]{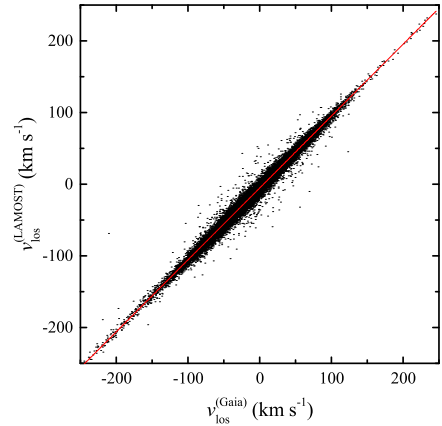}
}
 \caption{Line-of-sight velocities of the LAMOST DR8 K giants versus those provided by the {\it Gaia} EDR3. The red line denotes the linear fit with a slope of 1.0.}
\end{figure}\label{fig1}

\begin{figure}
\centering
\centerline{
\includegraphics[width=0.65\hsize]{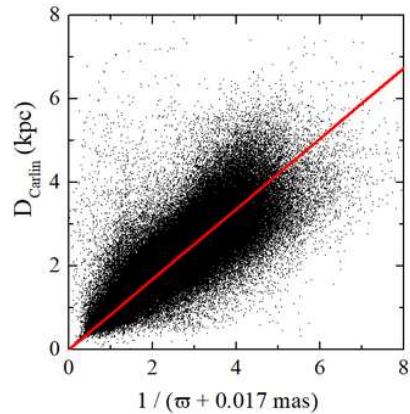}
}
 \caption{Distances derived by the method of \citet{Carlin15} versus the inverse of the {\it Gaia} parallaxes for the $\sim355000$ K giants with high-precision trigonometric parallaxes (i.e. $\sigma_\varpi/\varpi<0.1$). The red line denotes the linear fit with a slope of 0.842.}
\end{figure}\label{fig2}

\begin{figure}
\centering
\centerline{
\includegraphics[width=0.85\hsize]{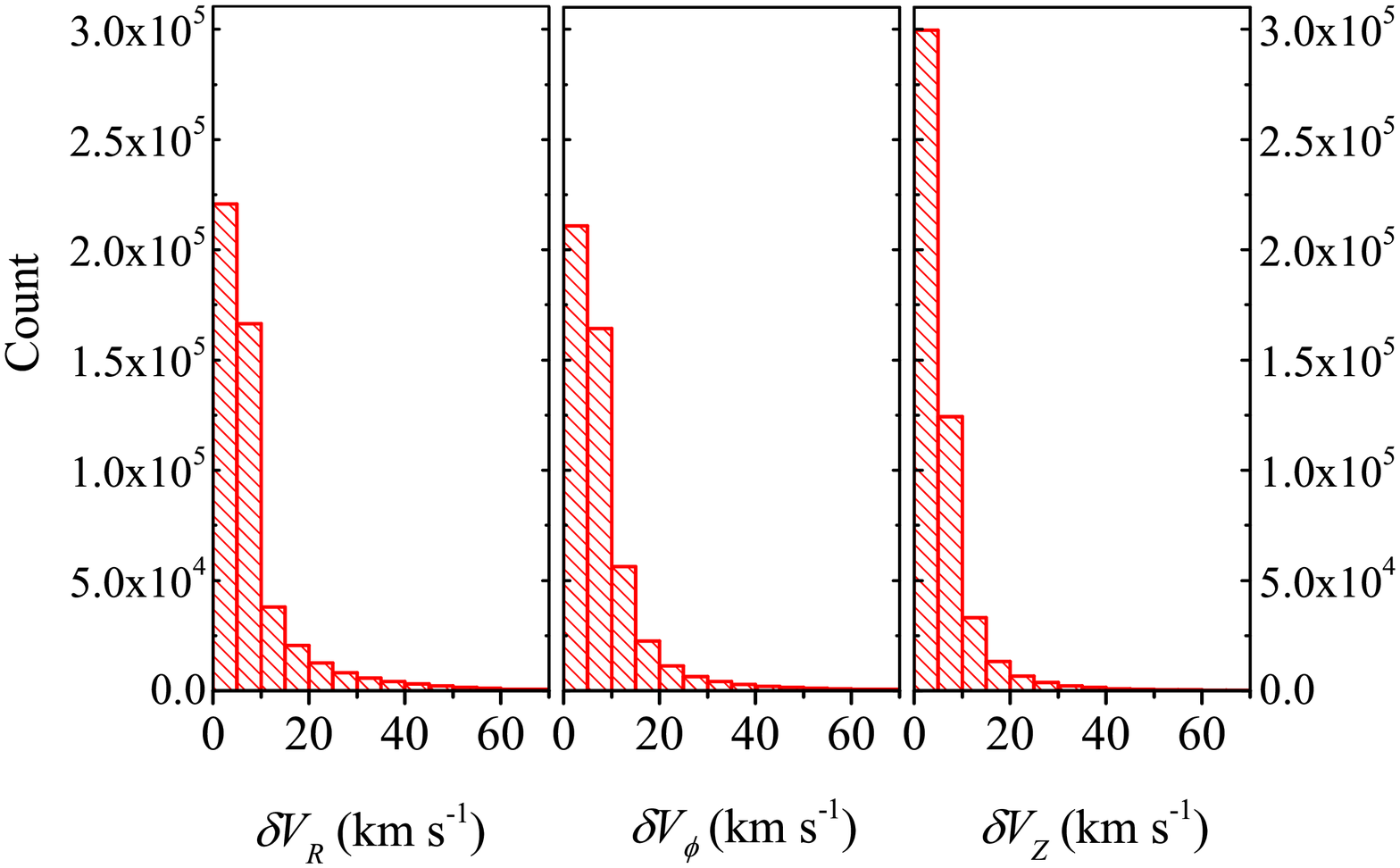}
}
 \caption{Error histograms for space velocities $V_R$, $V_\phi$ and $V_Z$ respectively for all the selected K giants.}
\end{figure}\label{fig3}

\begin{figure}
\centering
\centerline{
\includegraphics[width=0.95\hsize]{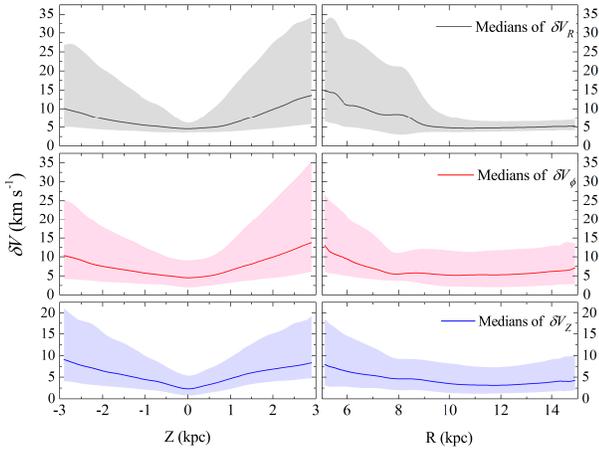}
}
 \caption{Velocity uncertainties as functions of $Z$ (left panels) and $R$ (right panels) respectively. The lines mark the medians of the uncertainties of $V_R$ (top panels), $V_\phi$ (middle panels) and $V_Z$ (bottom panels) respectively. The shaded regions, limited by the 16 and 84 percentiles, show areas enclosing 68 percent of the stars.}
\end{figure}\label{fig4}

\begin{figure}
\centering
\centerline{
\includegraphics[width=0.7\hsize]{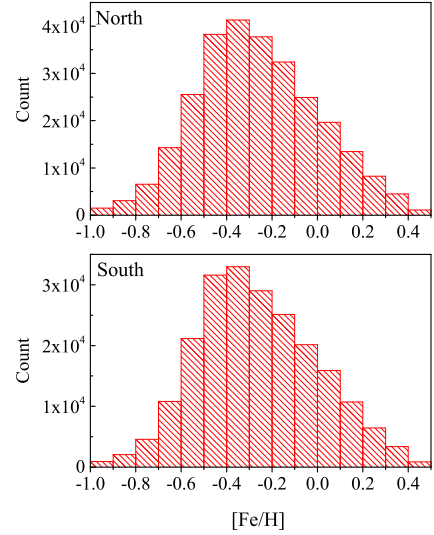}
}
 \caption{Histograms for the metallicity $\feh$ in the north ($Z>0$) and south ($Z<0$) disks respectively.}
\end{figure}\label{fig5}

In order to obtain a robust characterisation of stellar motions, we need information of proper motions and distances. The proper motions are obtained from the {\it Gaia} EDR3 (\citealt{Gaia Collaboration20}; \citealt{Lindegren20b}), which provides the best astrometric data nowadays. We adopt the criteria $\rm{ASTROMETRIC_{-}GOF_{-}AL}<3.0$ and $\rm{ASTROMETRIC_{-}EXCESS_{-}NOISE_{-}SIG}\leqslant2.0$ to select astrometrically well-behaved stars with good-fit statistic of the astrometric solution, coupled with a selection of the Renormalised Unit Weight Error (RUWE) $<1.4$ \citep{Lindegren20a} to favour stars with good astrometric quality. The distances, for the geometry, can be simply estimated by the inverse of the trigonometric parallax. However, such a method is feasible only for parallaxes with very high precision (\citealt{Bailer-Jones15}; \citealt{AB16}). Therefore, we employ the photometric distances derived from the measurements of distance modulus. To do this, we adopt the absolute magnitudes estimated using the method of \citet{Carlin15}, which is tailored specifically to the spectroscopically derived stellar parameters of LAMOST. Comparison with the apparent magnitude, which is obtained from the 2MASS point source catalog (\citealt{Cutri03}; \citealt{Skrutskie06}) in the $K_s$ band (2.16 $\mu$m), yields the distance modulus and then the distance. In order to correct for the extinction, we adopt $R(K_s)=0.306$ from \citet{Yuan13} as the reddening relative to $E(B-V)$ obtained from \citet{Schlegel98}. We obtain 545227 K giants with measurements of proper motions and distances.

We compare the derived photometric distances to the inverse of the {\it Gaia} parallax. For the comparison, we select $\sim355000$ {\it Gaia} stars with small relative parallax uncertainties $\sigma_\varpi/\varpi<0.1$ since using $1/\varpi$ as the distance estimate only works for stars with high precision. The comparison is shown in Fig. 2, in which the zeropoint of the {\it Gaia} parallax, i.e., 0.017 mas \citep{Lindegren20a}, has been added to the parallaxes. There is close agreement between the two distance scales, with a ratio of approximately 0.842 of the photometric distances to the inverse of parallaxes. In this case, we calibrate the photometric distances to more proper ones by dividing them by 0.842.

Throughout this paper we use Galactocentric cylindrical coordinates $(R,\,\phi,\,Z)$ with $R$ pointing outwards from the Galactic center (GC), the azimuthal $\phi$ in the direction of Galactic rotation, and $Z$ positive towards the north Galactic pole. The Galactocentric distance of the Sun is fixed at $R_0=8.122\,\kpc$ \citet{Gravity Collaboration18}, and the vertical offset of the Sun from the Galactic mid-plane is assumed to be $Z_\odot=20.8$ pc \citep{BB19}. The sample is located within $5<R<15\,\kpc$ and $|Z|<3\,\kpc$. Since the reference of the Galactocentric system is fixed at the GC, we need to know the velocity of the Sun. The azimuthal component of the solar velocity ($v_\odot$) can be obtained directly from the proper motion of Sgr A* along the Galactic longitude, i.e., $\mu_{\rm{Sgr A*}}=-6.411\,\masyr$ \citep{RB20}, which yields $v_\odot=246.8\,\kms$. The radial and vertical velocities of the Sun are taken from the widely used $U_\odot=11.1\,\kms$ and $W_\odot=7.25\,\kms$ given by \citet{Schonrich10}. In our sample, the velocity is confined to $|V_R|<200\,\kms$, $-80<V_\phi<320\,\kms$ and $|V_Z|<150\,\kms$ by using the $3\sigma$ criterion to avoid outlier velocities. Finally, a total number of $N=488209$ K giants remain.

The error histograms for the three-dimensional space velocities of our sample shown in Fig. 3 indicate good measurement qualities. The medians of $\delta V_R$, $\delta V_\phi$ and $\delta V_Z$ for all the selected K giants are 5.24, 5.65 and $3.97\,\kms$ respectively. Figure 4 gives the velocity uncertainties as functions of vertical distance and Galactocentic radius respectively. The measurement quality towards the Galactic anticenter is better than that towards the GC, which means that we can detailed detect stellar kinematics in the range of $|Z|<3\,\kpc$ from the inner disk near the solar circle to the distant outer disk.

\section{Results}

First and foremost, we construct the vertical pattern of stellar kinematics based on all the selected K giants as a function of Galactocentric radius. Then we compare the kinematics traced by populations with different metallicities. Figure 5 shows the histograms for the $\feh$ above and below the plane separately. When investigating the metallicity dependence of stellar motions, we divide the sample into 5 metallicity bins: the most metal-poor and metal-rich bins are $-1<\feh<-0.6$ and $0<\feh<0.5$ respectively; stars with intermediate metallicities are equally divided into bins with a width of 0.2 dex.

\subsection{Velocity moments in the radial and vertical directions}

The radial and vertical components of velocity moments, which carry certain kinematic details along the distance to the plane, play important roles in constraining the vertical structure of the disk. We use vertical bins of 0.2 kpc with the step of 0.1 kpc to investigate the velocity moments as functions of $Z$. Each bin contains no less than 30 stars. When making a comparison between different $\feh$, the bin width in $Z$ is 1 kpc and the step is 0.5 kpc. In each bin, we obtain the mean and dispersion of velocities in radial and vertical directions based on the Gaussian distribution (see Appendix).

\subsubsection{Radial bulk motions}
Figure 6 shows the mean of $V_R$, $\langle V_R\rangle$, as functions of $Z$ for populations at different $R$. For stars near the plane, the $\langle V_R\rangle$ changes from positive to negative when we move across the solar circle and back to positive beyond $R=10\,\kpc$. At higher disk there are north-south asymmetries in the $\langle V_R\rangle$ distribution, especially in the range of $9<R<13\,\kpc$ where the $\langle V_R\rangle$-$Z$ pattern relates to an outward flow with velocities up to around $20\,\kms$ at $Z<0$ and alternate outward and inward flows with velocities $\lesssim10\,\kms$ at $Z>0$. The $\langle V_R\rangle$ increases with the increasing of $|Z|$ within $-2<Z<0$, with a gradient of $\sim4-7\,\kmskpc$ at $5<R<11\,\kpc$ and $\sim0-5\,\kmskpc$ at $11<R<15\,\kpc$. In the north disk, the $\langle V_R\rangle$-$Z$ gradient for $R<10\,\kpc$ is positive, shallowing as $R$ increases, while for $10<R<13\,\kpc$ the gradient stays positive only between $z\sim0$ and $0.5\,\kpc$, and reverses beyond $Z=0.5\,\kpc$. The $Z$-dependence of $\langle V_R\rangle$ is nearly vanished for the radial slice farthest from the GC.

\begin{figure*}
\centering
\centerline{
\includegraphics[width=0.99\hsize]{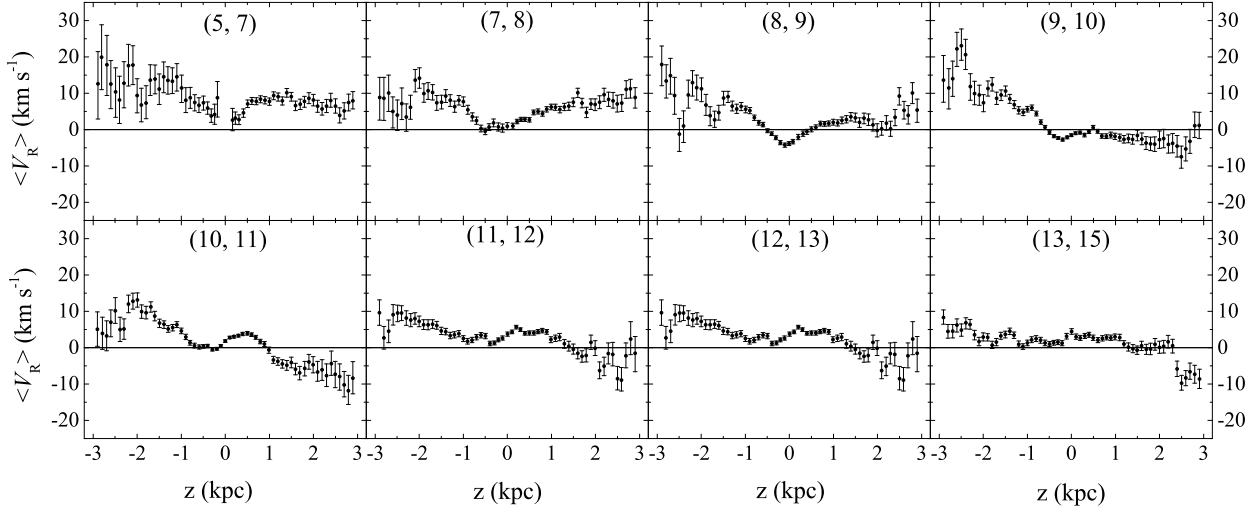}
}
 \caption{Mean radial velocities as functions of the vertical distance to the Galactic mid-plane, coupled with the measurement errors. The horizontal axis is the mean vertical distance of each vertical bin. The range of the Galactocentric radius, in unit of kpc, is labeled at the top of each panel.}
\end{figure*}\label{fig6}

\begin{figure*}
\centering
\centerline{
\includegraphics[width=0.99\hsize]{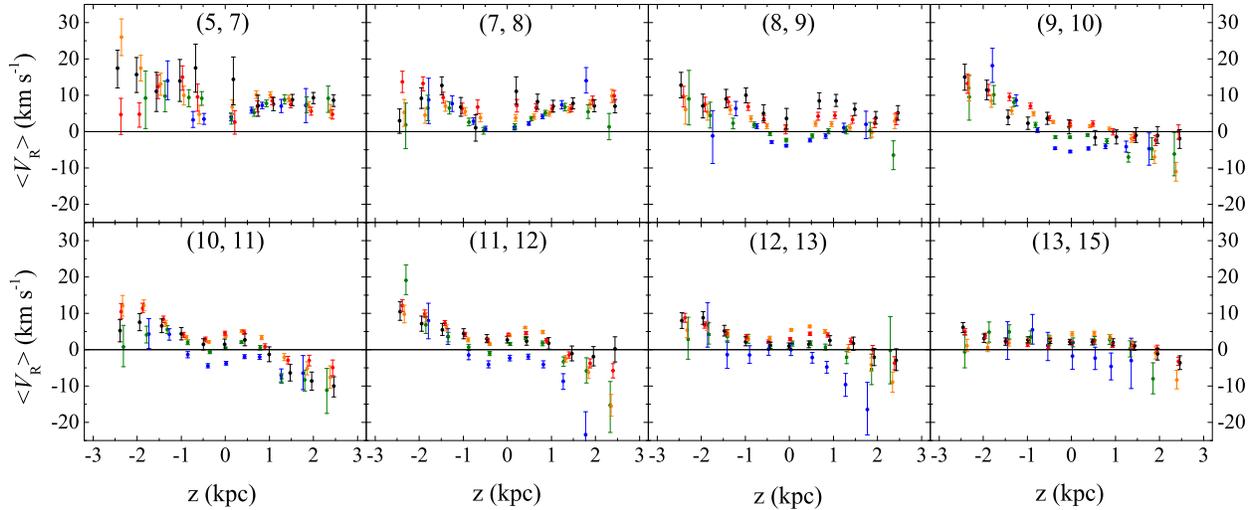}
}
 \caption{Mean radial velocities as functions of the vertical distance to the Galactic mid-plane for different metallicities, coupled with the measurement errors. The black, red, orange, olive and blue dots denote the metallicity range $-1<\feh<-0.6$, $-0.6<\feh<-0.4$, $-0.4<\feh<-0.2$, $-0.2<\feh<0$ and $0<\feh<0.5$, respectively. The horizontal axis is the mean vertical distance of each vertical bin. The range of the Galactocentric radius, in unit of kpc, is labeled at the top of each panel.}
\end{figure*}\label{fig7}

Figure 7 presents the comparison between different $\feh$. There is no significant relation between $\langle V_R\rangle$ and $\feh$ in the inner disk. In the outer disk within $8<R<12\,\kpc$, metal-poor populations have higher $\langle V_R\rangle$ than metal-rich ones, though the vertical behaviors of $\langle V_R\rangle$ for different $\feh$ are qualitatively consistent. The gap of $\langle V_R\rangle$ between metallicities tends to narrow when $|Z|\gtrsim1\,\kpc$ and $R>12\,\kpc$, indicating a mixture of populations at high disk and large Galactocentric radius.

\begin{figure*}
\centering
\centerline{
\includegraphics[width=0.99\hsize]{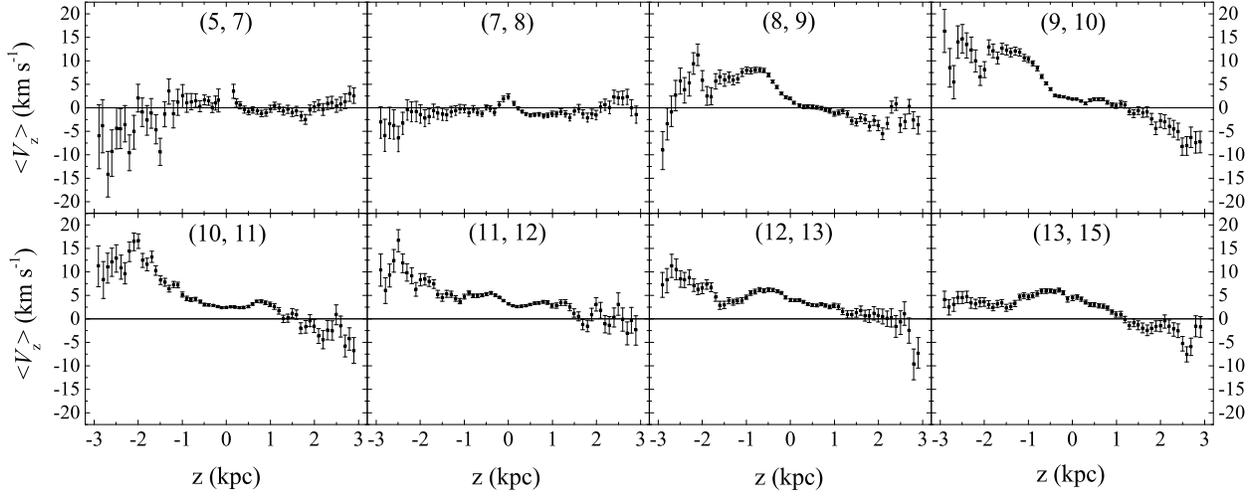}
}
 \caption{As in Fig. 6, but for the mean vertical velocities.}
\end{figure*}\label{fig8}

\begin{figure}
\centering
\centerline{
\includegraphics[width=1.0\hsize]{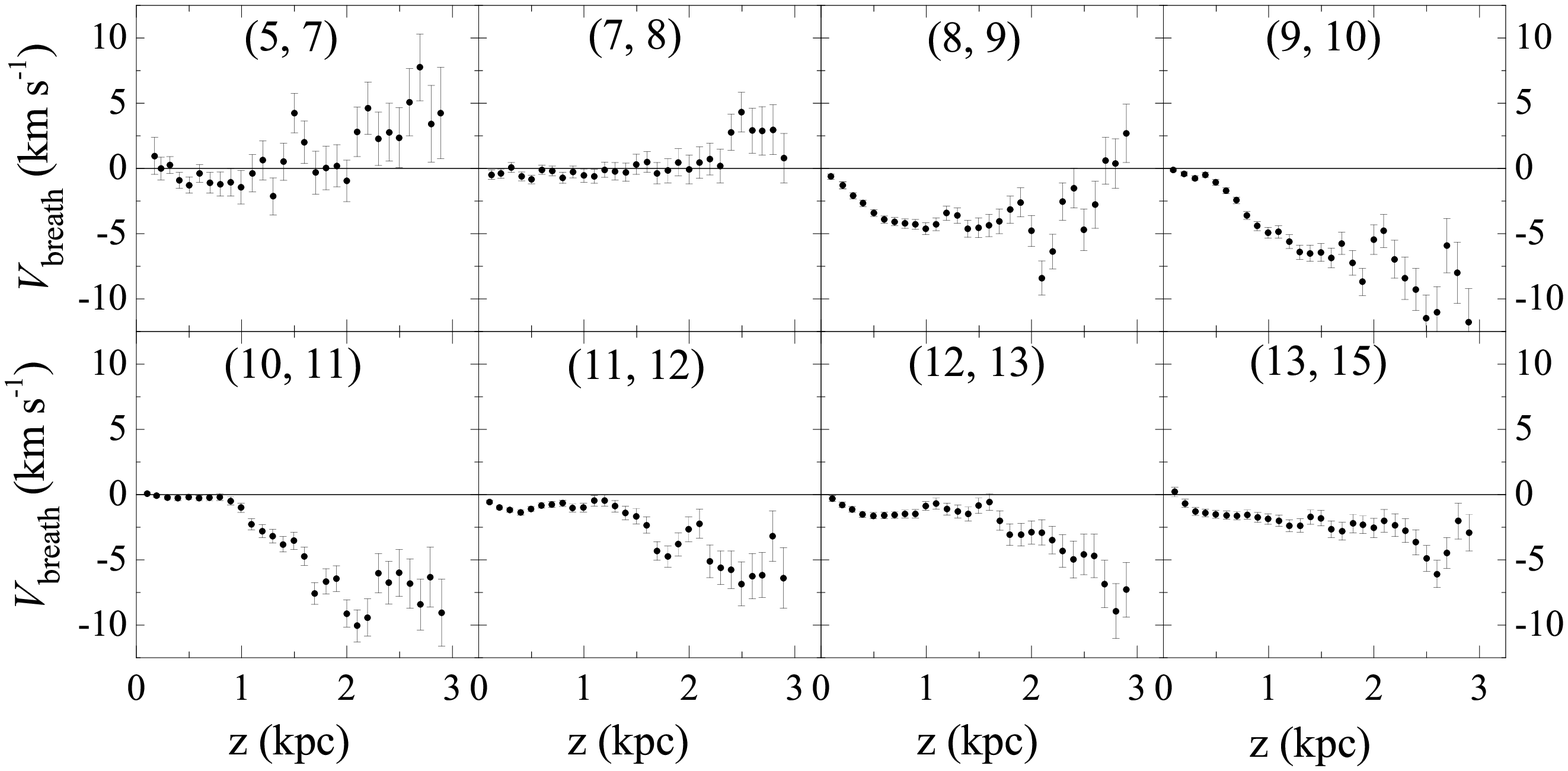}
}
 \caption{Breathing velocities as functions of the height from the Galactic mid-plane. The range of the Galactocentric radius, in unit of kpc, is labeled at the top of each panel.}
\end{figure}\label{fig9}

\begin{figure}
\centering
\centerline{
\includegraphics[width=1.0\hsize]{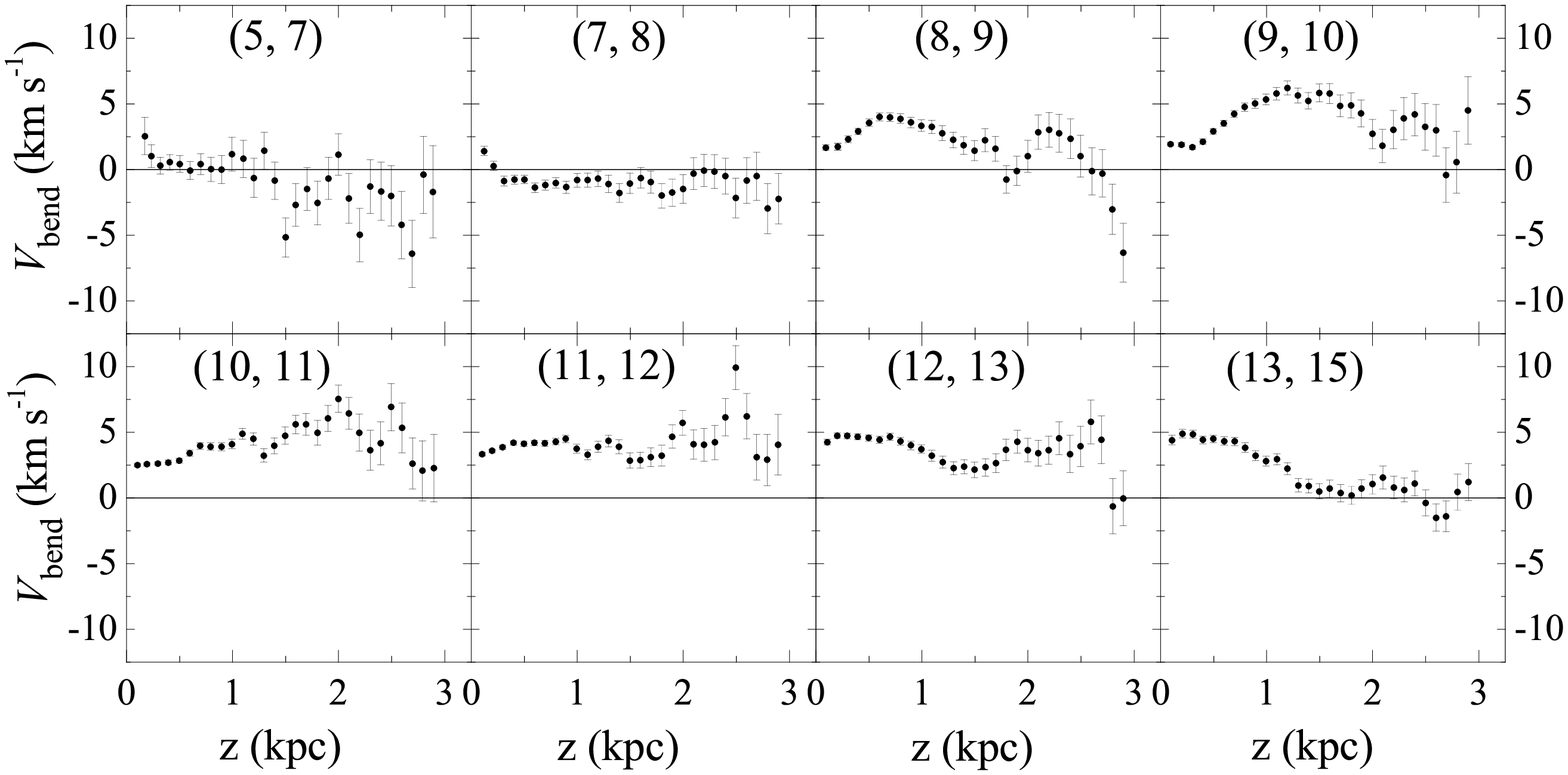}
}
 \caption{As in Fig. 9, but for the bending velocities.}
\end{figure}\label{fig10}

\begin{figure}
\centering
\centerline{
\includegraphics[width=1.0\hsize]{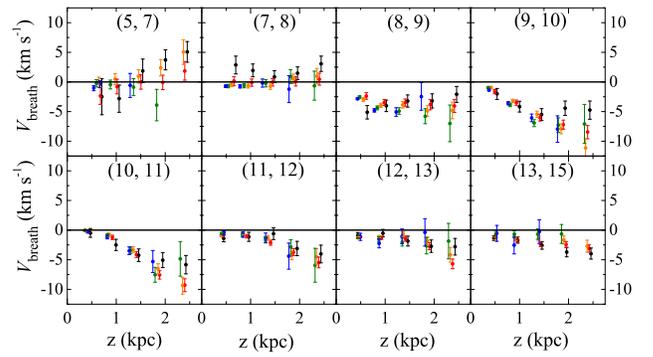}
}
 \caption{Breathing velocities as functions of the height from the Galactic mid-plane for different metallicities. The black, red, orange, olive and blue dots denote the metallicity range $-1<\feh<-0.6$, $-0.6<\feh<-0.4$, $-0.4<\feh<-0.2$, $-0.2<\feh<0$ and $0<\feh<0.5$, respectively. The range of the Galactocentric radius, in unit of kpc, is labeled at the top of each panel.}
\end{figure}\label{fig11}

\begin{figure}
\centering
\centerline{
\includegraphics[width=1.0\hsize]{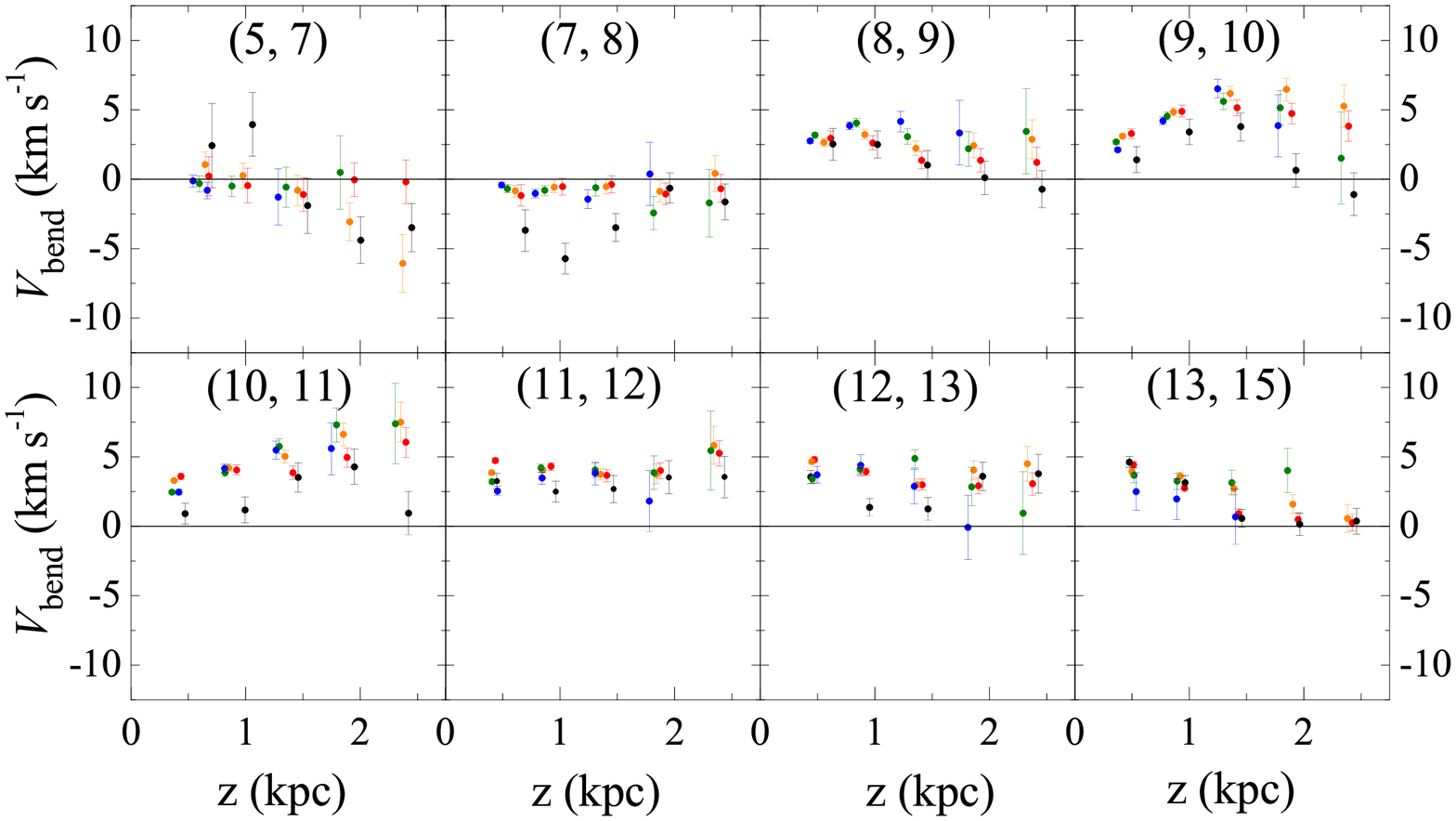}
}
 \caption{As in Fig. 11, but for the bending velocities.}
\end{figure}\label{fig12}

\begin{figure*}
\centering
\centerline{
\includegraphics[width=0.99\hsize]{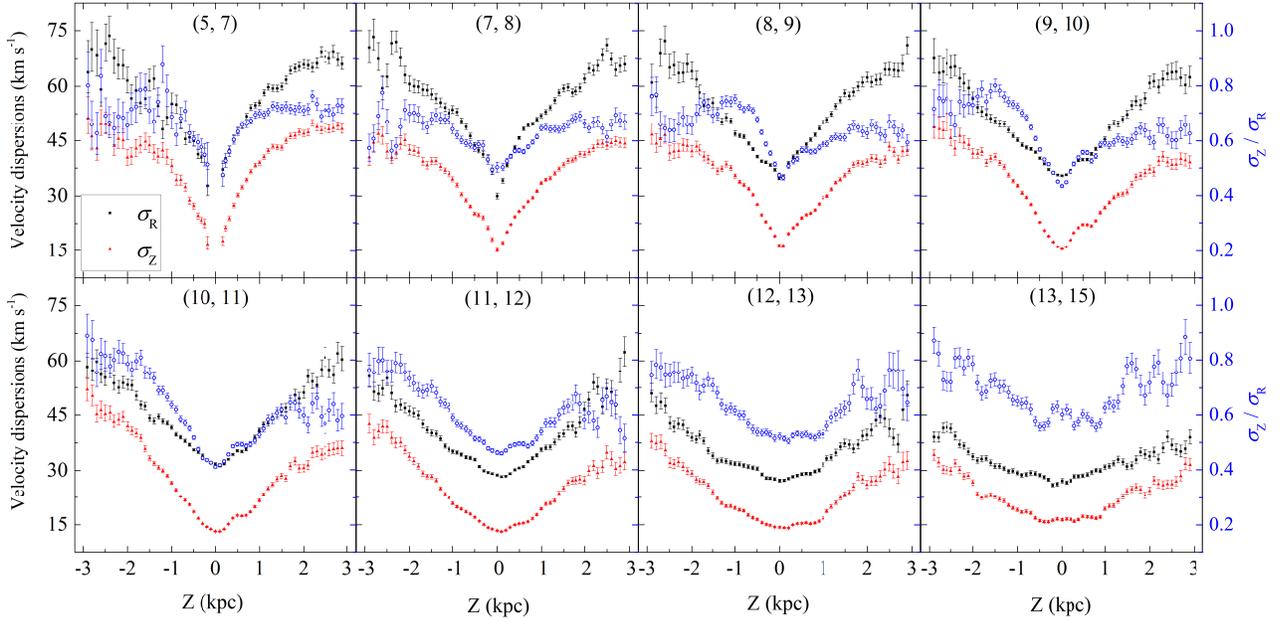}
}
 \caption{As Fig. 6, but for the radial (black squares) and vertical (red triangles) velocity dispersions, associated with the ratio of vertical to radial dispersions (blue circles).}
\end{figure*}\label{fig13}

\begin{figure*}
\centering
\centerline{
\includegraphics[width=0.99\hsize]{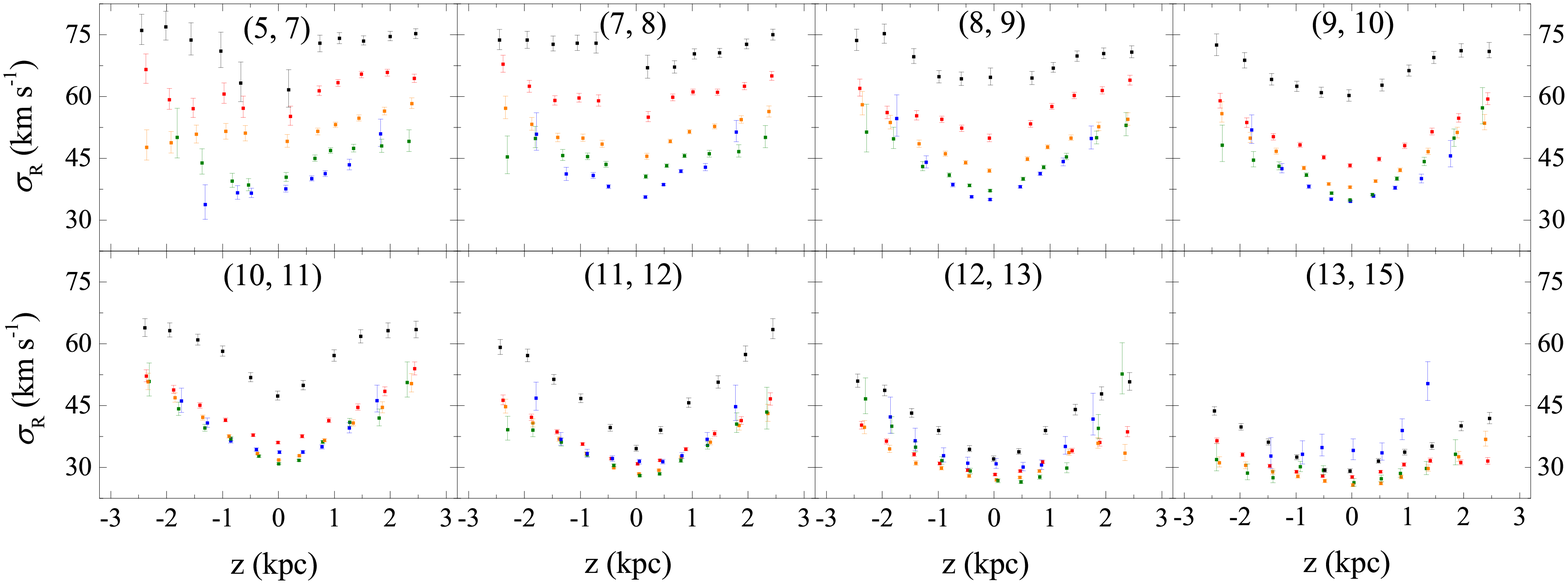}
}
 \caption{As Fig. 7, but for the radial velocity dispersions.}
\end{figure*}\label{fig14}

\begin{figure*}
\centering
\centerline{
\includegraphics[width=0.99\hsize]{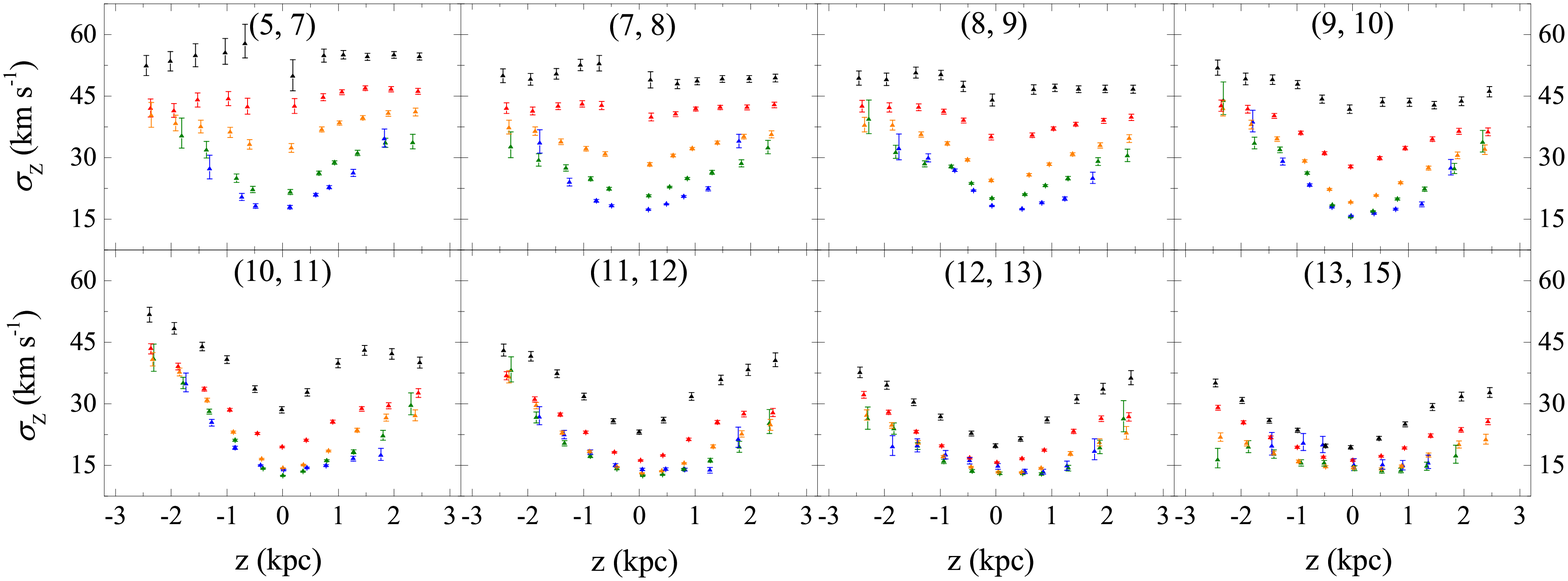}
}
 \caption{As Fig. 7, but for the vertical velocity dispersions.}
\end{figure*}\label{fig15}

\subsubsection{Breathing and bending motions}
The vertical patterns of the mean vertical velocities $\langle V_Z\rangle$ given by Fig. 8 show a north-south asymmetry, which is most notable in the range $8<R<12\,\kpc$. Figures 9 and 10 give the breathing and bending velocities as functions of $|Z|$ respectively, which are calculated based on the the definition in \citet{Gaia Collaboration18}, i.e.,
\setlength{\mathindent}{0pt}
\begin{eqnarray}
V_{\rm{breath}}&=&\frac{1}{2}(\langle V_Z\rangle^{\rm{North}}-\langle V_Z\rangle^{\rm{South}}),\nonumber\\
V_{\rm{bend}}&=&\frac{1}{2}(\langle V_Z\rangle^{\rm{North}}+\langle V_Z\rangle^{\rm{South}}).
\end{eqnarray}\label{Eq1}

In the inner disk, there is no significant breathing motion at $|Z|\lesssim2\,\kpc$; beyond 2 kpc we find positive $\vbreath$ as large as around $5\,\kms$, suggesting a rarefaction effect at high disk. The phase of the breathing mode reverses outside the solar circle: the vertical motion is dominated by a contraction movement, whose amplitude shows dependence on both $|Z|$ and $R$: at $|Z|\sim1\,\kpc$, the $|\vbreath|$ is $\sim5\,\kms$ within $8<R<10\,\kpc$ while stays $\sim0-2\,\kms$ within $10<R<15\,\kpc$; at $|Z|\sim2\,\kpc$, the $|\vbreath|$ increases with $R$ when $R<11\,\kpc$ while the trend reverses beyond $R=11\,\kpc$. In the range between $R=9$ and 15 kpc, the amplitude of the breathing motion increases towards higher disk, the gradient of $|\vbreath|$ with $|Z|$ decreasing from $\sim3.7\,\kmskpc$ at $9<R<10\,\kpc$ to $\sim1.4\,\kmskpc$ at $13<R<15\,\kpc$.

Similar to the breathing velocity, the sign of the bending velocity reverses when we move across the solar circle (see Fig. 10). In the inner disk, the bending motion is trivial. The mean of $\vbend$ is around $-1\,\kms$ at $R<8\,\kpc$. In the outer disk, the $\vbend$ stays positive (up to $\sim5\,\kms$) within $|Z|<2.5\,\kpc$, suggesting a bending movement towards the north. There is a flat or positive $\vbend$-$|Z|$ gradient at $|Z|<1\,\kpc$. When we move to $|Z|>1\,\kpc$, a negative $\vbend$-$|Z|$ gradient dominates the bending mode pattern.

Figures 11-12 present the breathing and bending velocities for populations with different metallicities. The differences between the amplitudes of the breathing motions for different $\feh$ are only evident in the solar vicinity, where the $|\vbreath|$ increases with the increasing of $\feh$. In the region far away from the Sun, the $|\vbreath|$ for different $\feh$ are nearly consistent. As for the bending motion, we find a positive $\vbend$-$\feh$ trend at $8<R<9\,\kpc$, similar to the metallicity trend of $|\vbreath|$. Outside of this slice, the $\vbend$ is almost independent on $\feh$, except for that in the range of $7<R<12\,\kpc$ the $\vbend$ of the most metal-poor population is lower than that of other populations.

\subsubsection{Velocity dispersions in the radial and vertical directions}
The radial and vertical velocity dispersions are presented in Fig. 13. Both $\sigma_R$ and $\sigma_Z$ increase with the increasing distance to the plane, which means that populations further away from the plane have been stronger affected by the heating mechanisms so that been diffused through a larger fraction of phase space. The radial dispersions on both sides of the disk are approximately comparable, only with small differences at $R<10\,\kpc$ and $|Z|<2\,\kpc$ that the $\sigma_R$ above the plane is on average larger than that below the plane by no more than a few $\kms$. A more significant north-south asymmetry exists in $\sigma_Z$ within $7<R<13\,\kpc$ such that $\sigma_Z$ of $Z<0$ stars are larger than that of $Z>0$ stars by up to $\sim15\,\kms$.

The velocity dispersions decrease towards the outer disk. In comparison, there is no pronounced trend between $\sigma_Z/\sigma_R$ and $R$ (see Fig. 13). The mean $\sigma_Z/\sigma_R$ is 0.49 at $Z\sim0$, then increases with the increasing $|Z|$, notably when $|Z|\lesssim1.5\,\kpc$. The mean $\sigma_Z/\sigma_R$ is 0.73 at $Z<-1.5\,\kpc$ and 0.67 at $Z>1.5\,\kpc$.

The metallicity dependence of $\sigma_R$ and $\sigma_Z$ is shown in Figs. 14 and 15 respectively. The gradient of the dispersions with respect to $|Z|$ is larger for populations with higher metallicities. The increasing of both $\sigma_R$ and $\sigma_Z$ with the decreasing $\feh$ is notable within $5<R<10\,\kpc$, while when we move towards the Galactic anticenter, the dependence of dispersions on $\feh$ weakens, especially for the radial component.

\subsubsection{Tilt of the velocity ellipsoid}\label{sec3.1.4}
We present the tilt angles of the velocity ellipsoid in Fig. 16, which are calculated by the covariance between $V_R$ and $V_Z$ via $\tilt\equiv\frac{1}{2}\arctan[2\sigma_{RZ}/(\sigma_R^2-\sigma_Z^2)]$. The panels of $5<R<9\,\kpc$ show a clear positive trend between $\tilt$ and $Z$. The vertical patterns of $\tilt$ can be fitted well by $\tilt=(0.827\pm0.040)\arctan(Z/R)-(0.007\pm0.011)$, $\tilt=(0.793\pm0.034)\arctan(Z/R)-(0.0032\pm0.0061)$ and $\tilt=(0.520\pm0.041)\arctan(Z/R)+(0.0370\pm0.0064)$ at $5<R<7\,\kpc$, $7<R<8\,\kpc$ and $8<R<9\,\kpc$, respectively, indicating that the orientation of the velocity ellipsoid is near-spherical in the inner disk and turns more horizontal when we move across the solar circle.

\begin{figure*}
\centering
\centerline{
\includegraphics[width=0.99\hsize]{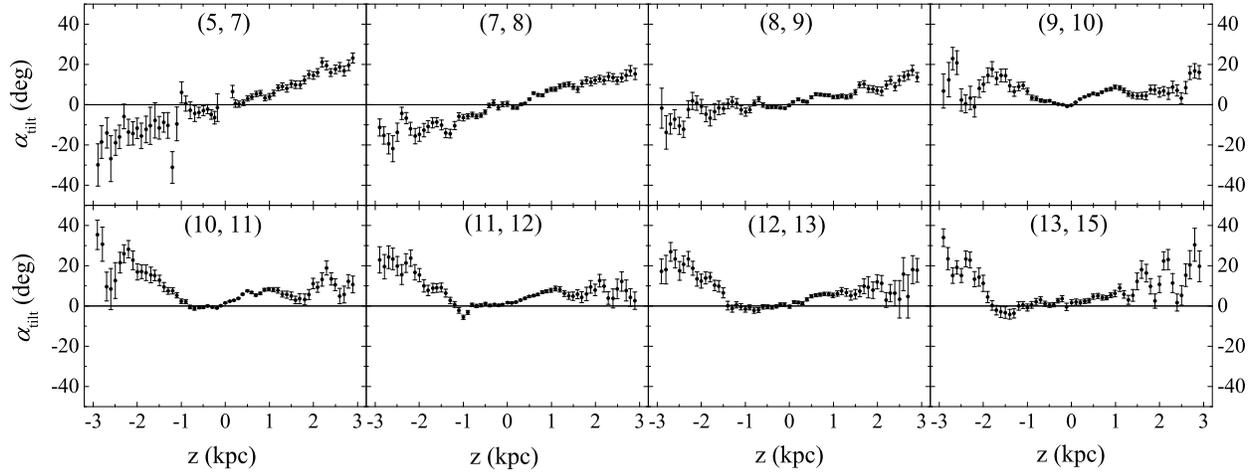}
}
 \caption{As Fig. 6, but for the tilt angles.}
\end{figure*}\label{fig16}

In the range of $9<R<15\,\kpc$, the $\tilt$-$\arctan(Z/R)$ gradient ranges between $\sim0.4$ and $\sim1.2$ for $-1<Z<3\,\kpc$. It is interesting to find that the vertical behavior of $\tilt$ reverses at $Z\lesssim-1\,\kpc$, which breaks the antisymmetry in the $\tilt$-$Z$ plane. Figure 17 shows the tilt angles for different metallicities. The vertical patterns of $\tilt$ for different $\feh$ are basically consistent.

\begin{figure*}
\centering
\centerline{
\includegraphics[width=0.99\hsize]{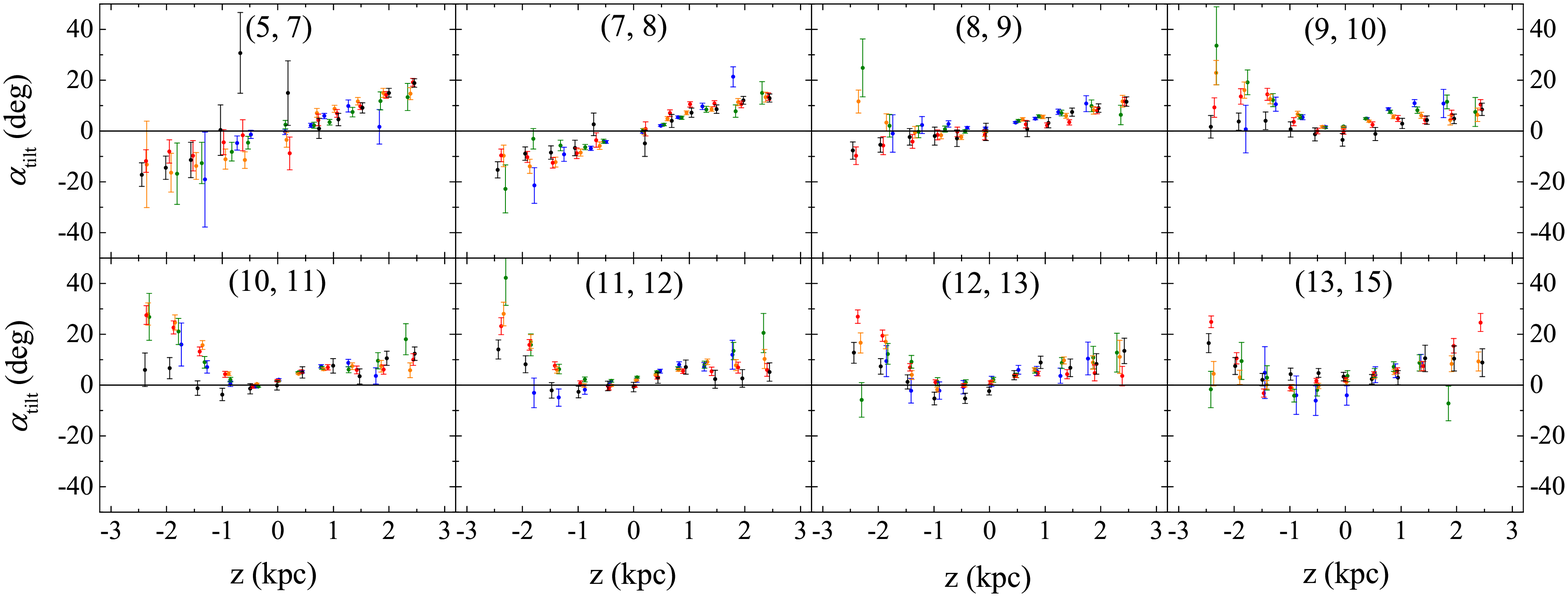}
}
 \caption{As Fig. 7, but for the tilt angles.}
\end{figure*}\label{fig17}

\subsection{Vertical structures of azimuthal velocities}

In this subsection, we focus on the azimuthal velocities as functions of distance to the plane. We take a bin of 0.5 kpc and a step of 0.25 kpc to construct the vertical structure of velocity moments in the azimuthal direction. Each bin contains no less than 100 stars. We employ a non-Gaussian distribution function to fit the observed distribution of the azimuthal velocity (see Appendix) and obtain the first and second moments from the distribution function.

\begin{figure*}
\centering
\centerline{
\includegraphics[width=0.9\hsize]{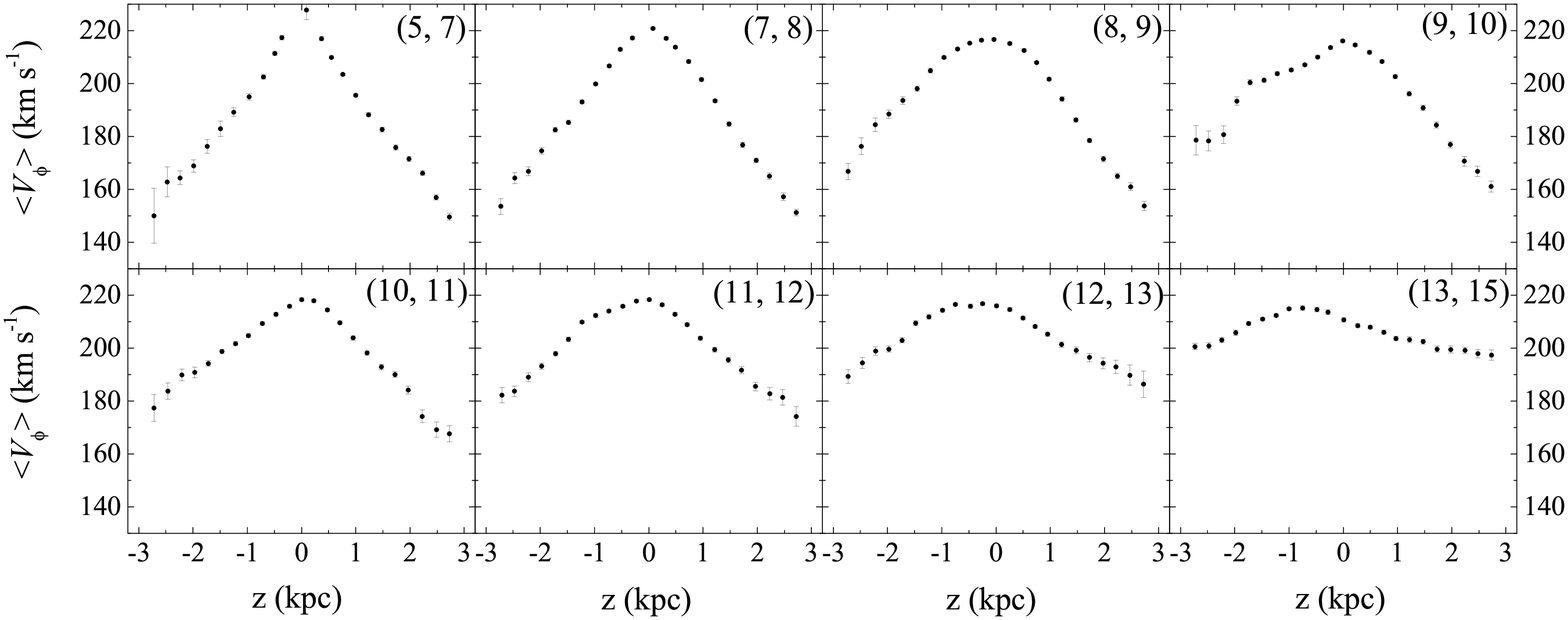}
}
 \caption{As Fig. 6, but for the mean azimuthal velocities.}
\end{figure*}\label{fig18}

\begin{figure*}
\centering
\centerline{
\includegraphics[width=0.9\hsize]{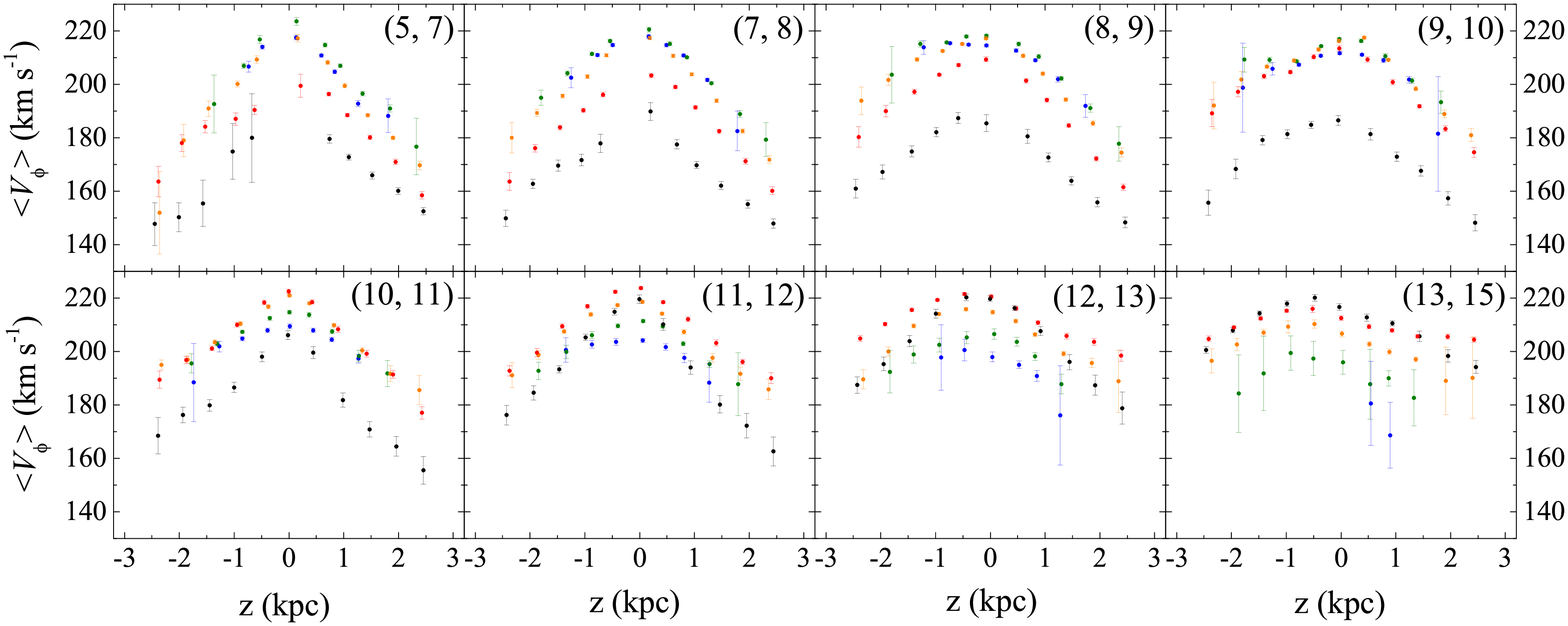}
}
 \caption{As Fig. 7, but for the mean azimuthal velocities.}
\end{figure*}\label{fig19}

Figure 18 shows the mean azimuthal velocity, $\langle V_\phi\rangle$, as functions of $Z$ in different radial slices. The $\langle V_\phi\rangle$ at $Z\sim0$ shows no significant radial trend, suggesting a near-flat rotation curve in the plane. The $\langle V_\phi\rangle$ decreases as we move away from the plane. Table 1 gives the absolute values of the vertical gradient in $\langle V_\phi\rangle$ above and and below the plane respectively. The $|d\langle V_\phi\rangle/dZ|$ in the south disk is smaller than that in the north, especially in the range of $5<R<12\,\kpc$, suggesting that stars below the plane are on average rotating faster than stars above. Figure 19 presents the $\langle V_\phi\rangle$ patterns for different populations. Generally, the azimuthal velocities of populations with $-1<\feh<-0.4$ increase towards the outer disk, while the trend reverses for $-0.2<\feh<0.5$.

\begin{figure*}
\centering
\centerline{
\includegraphics[width=0.9\hsize]{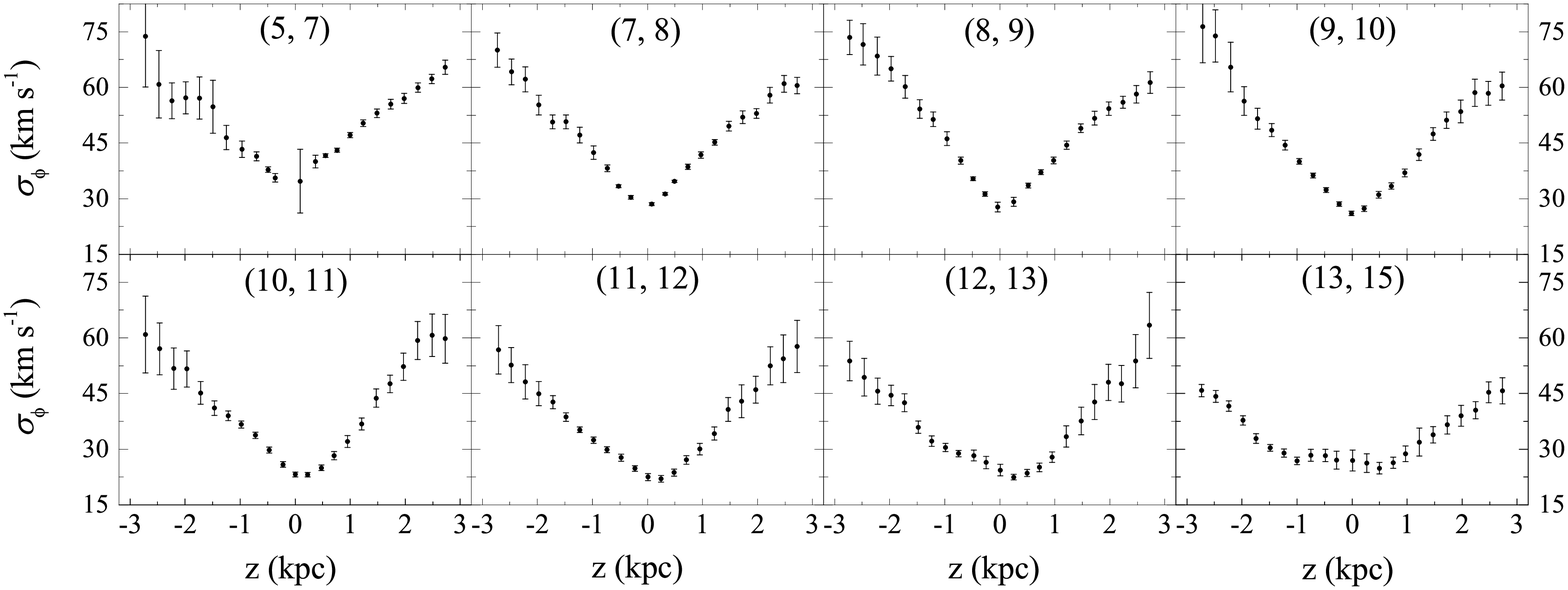}
}
 \caption{As Fig. 6, but for the azimuthal velocity dispersions.}
\end{figure*}\label{fig20}

\begin{figure*}
\centering
\centerline{
\includegraphics[width=0.9\hsize]{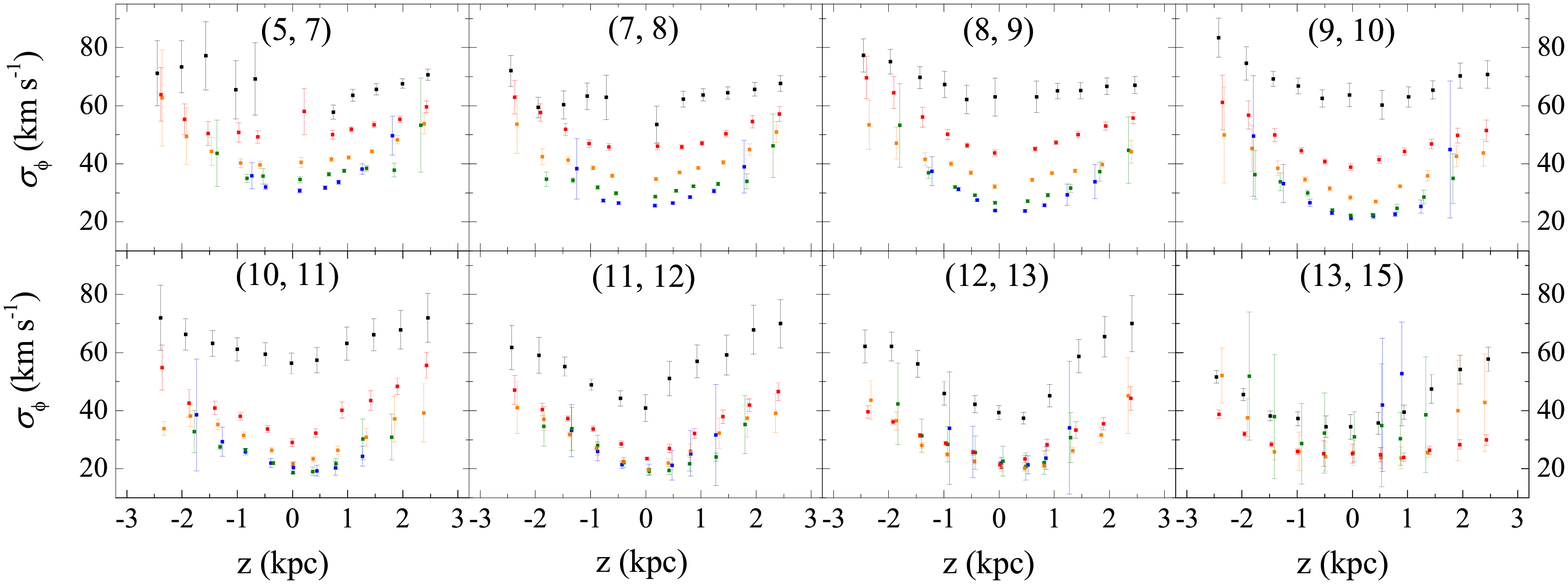}
}
 \caption{As Fig. 7, but for the azimuthal velocity dispersions.}
\end{figure*}\label{fig21}

\begin{table*}
\newcommand{\thirdsmall}{\fontsize{8.2pt}{\baselineskip}\selectfont}
\renewcommand\arraystretch{1}
\newcommand{\ba}[1]{\raisebox{-1.2ex}[0pt][0pt]{\shortstack{#1}}}
\newcommand{\bb}[1]{\raisebox{-3ex}[0pt][0pt]{\shortstack{#1}}}
\caption{Absolute values of the vertical gradient in the mean azimuthal velocities in the north ($Z>0$) and south ($Z<0$) disks respectively within different radius slices.}
\thirdsmall
\begin{tabular}{cccccccccc}
\hline\hline
 $R\,(\kpc)$ &  &$(5,\,7)$ &$(7,\,8)$ &$(8,\,9)$ &$(9,\,10)$ &$(10,\,11)$ &$(11,\,12)$ &$(12,\,13)$ &$(13,\,15)$\\
\hline
\ba{$|d\langle V_\phi\rangle/dZ|\,(\kmskpc)$} &$Z>0$ &$28.21\pm0.79$ &$27.64\pm0.57$ &$24.9\pm1.0$ &$21.59\pm0.79$ &$20.35\pm0.73$ &$16.45\pm0.39$ &$11.00\pm0.29$ &$4.92\pm0.30$\\
&$Z<0$ &$26.79\pm0.84$ &$24.74\pm0.74$ &$18.6\pm1.2$ &$14.4\pm1.3$ &$14.49\pm0.40$ &$14.73\pm0.98$ &$10.4\pm1.0$ &$5.21\pm0.98$\\
\hline
\end{tabular}
\end{table*}

The vertical patterns of the azimuthal dispersion $\sigma_\phi$ and their dependence on $\feh$ are given by Figs. 19 and 20 respectively. As expected, the $\sigma_\phi$ increases towards higher disk and lower metallicity as a whole. The nouth-south asymmetry in $\sigma_\phi$ is less pronounced than that in $\langle V_\phi\rangle$. In the innermost slice, the $\sigma_\phi$ is asymmetric within $|Z|\lesssim1\,\kpc$, with larger values for north disk stars. The reverse applies for the $\sigma_\phi$ pattern of the the outer disk.

\section{An implication of the vertical structure of stellar kinematics: the flare of the Galactic disk}

The vertical structure of stellar kinematics gives us an insight into the stellar distribution in the vertical direction. In this section, we use the observed velocity ellipsoid shown in Sect. 3 to find the disk scaleheights $h_z$ at different $R$, which allows us to detect the flaring strength of the disk.

We employ the method from \citet{MoniBidin12} and \citet{Lopez-Corredoira20} to estimate the $h_z$ and its radial behavior. The basic idea is to fit the expression for the surface density $\Sigma(Z)$ derived from the Jeans equations to the measured velocity dispersions and covariance:
\begin{small}
\begin{eqnarray}
 \Sigma(Z)&=&\frac{1}{2\pi G}\Big[k_1\cdot\int_0^Z\sigma_R^2\rm{d}Z+k_2\cdot\int_0^Z\sigma_\phi^2\rm{d}Z+k_3\cdot\sigma_{RZ} \nonumber\\
 &&+\frac{\sigma_Z^2}{h_z}-\frac{\partial \sigma_Z^2}{\partial Z}+|Z|\sigma_{RZ}\frac{\partial}{\partial R}\Big(\frac{1}{h_z}\Big)+\int_0^Z|Z|\frac{\sigma_R^2}{R}\frac{\partial}{\partial R}\Big(\frac{1}{h_z}\Big)\rm{d}Z \nonumber\\
 &&+\int_0^Z|Z|\sigma_R^2\frac{\partial^2}{\partial R^2}\Big(\frac{1}{h_z}\Big)\rm{d}Z\Big] \nonumber\\
 &=&2\rho(R,\,Z=0)\cdot h_z(R)\cdot \big[1-e^{-Z/h_z(R)}\big].
\end{eqnarray}\label{Eq2}
\end{small}
The constants $k_1$, $k_2$ and $k_3$ are defined by
\begin{eqnarray}
k_1&=&\frac{3}{R_0\cdot h_R}-\frac{2}{h_R^2},\nonumber\\
k_2&=&-\frac{1}{R_0\cdot h_R},\nonumber\\
k_3&=&\frac{3}{h_R}-\frac{2}{R_0},
\end{eqnarray}\label{Eq3}
where $h_R$ is the radial scalelength. In this work we fix $h_R\equiv4.9\,\kpc$ from the result of \citet{Lopez-Corredoira20}.

We derive the fitting values for $h_z$ using an iteration method. Using the initial values of $h_z(R)$, the corresponding derivatives of $1/h_z$, namely $\frac{\partial}{\partial R}\Big(\frac{1}{h_z}\Big)$ and $\frac{\partial^2}{\partial R^2}\Big(\frac{1}{h_z}\Big)$, and the $\Sigma(Z)$ can be derived for every radial slice. Then the new values of $h_z(R)$ are determined from Eq. (2) using the least square method. We adopt the expression $h_z(R)=0.533+0.103(R-R_0)$ from \citet{Lopez-Corredoira20} to calculate the initial values for $h_z$. The iteration procedure is repeated until convergence is achieved.

When using Eq. (2) to estimate $h_z$ we adopt two assumptions behind the Jeans equations, namely the stationary of the disk and the axisymmetry of the system. There is evidence that some non-axisymmetries exist in the disk, which means that the Jeans equations do not strictly describe the Galactic dynamics and the $h_z(R)$ derived from Eq. (2) is an approximation of the scaleheight for the observation range.

Figure 22 shows the estimated $h_z$ derived from the measured velocity ellipsoids in the north and south disks respectively. (For $Z<0$ we use $|Z|$ instead of $Z$ and $-\sigma_{RZ}$ instead of $\sigma_{RZ}$ in Eq. (2).) It is interesting to find that the $h_z$ traced by south disk stars is larger than that traced by north disk stars, with difference decreasing with the increasing $R$. Nonetheless, the flaring features in the north and south disks are similar, in which the noticeable flaring begins at $R\sim8-9\,\kpc$. We use a second-order polynomial fit to constrain the flaring. The new expressions for $h_z(R)$ are $h_z=(0.265\pm0.031)+(0.100\pm0.025)(R-R_0)+(0.0261\pm0.0076)(R-R_0)^2\,\kpc$ for the north disk, and $h_z=(0.471\pm0.029)+(0.091\pm0.023)(R-R_0)+(0.0186\pm0.0069)(R-R_0)^2\,\kpc$ for the south disk.

\begin{figure}
\centering
\centerline{
\includegraphics[width=0.85\hsize]{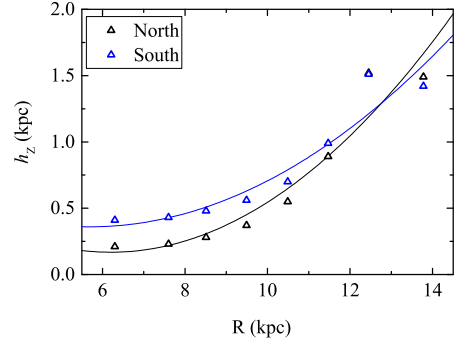}
}
 \caption{Scale heights as functions of the Galactocentric radius in the north (black squares) and south (blue triangles) disks respectively. The fits of the scale heights are marked by the black and blue lines respectively.}
\end{figure}\label{fig22}

\section{Discussion}

Due to high luminosities and a large sample size, the K giant stars are increasingly being used as tracers to probe Galactic kinematics and dynamics over a large section of the Milky Way. The huge number spectra observed by the LAMOST survey yield an important K giants sample with stellar parameters and distance estimates. With the help of the high-precision astrometric data provided by the {\it Gaia} EDR3, we can characterize the stellar disk  in a wider range of vertical distance with unprecedented precision. We find north-south asymmetries in all the three components of mean velocities, where the vertical patterns of the radial and vertical components are more complex than the azimuthal components. The non-zero $\langle V_R\rangle$ and $\langle V_Z\rangle$ and their variations with $Z$ in different radial slices suggest that there exist perturbations with internal or external origins acting on stellar motions.

We find non-zero mean radial velocities ranging from $\sim-10\,\kms$ to $\sim20\,\kms$, which are dependent on the distance to the plane. The $\langle V_R\rangle$ decreases towards larger $R$ as a whole, coupled with shallowing vertical gradients. Considering different metallicity bins, there is evidence that the $\langle V_R\rangle$ of metal-rich stars decreases with $R$ more steeply than that of metal-poor stars in the range of $8<R<12\,\kpc$, which is in agreement with the finding of \citet{Wojno18} from the RAVE FGK turn-off stars. The observed radial flow and its alternate direction near the plane are likely generated by the wave-like spiral structures in the plane (\citealt{Faure14}; \citealt{Monari16}). \citet{Faure14} argued that the effect of the spirals on the stellar radial velocities keeps strong up to $Z\sim\pm0.5\,\kpc$. Besides, \citet{Monari14} found that the central bar could produce radial bulk motions within $R<10\,\kpc$ and give rise to a negative radial gradient in the radial bulk velocities in the range from $Z=0$ to $Z\sim2\,\kpc$. Within $8<R<12\,\kpc$, populations with different $\feh$ respond differently to the perturbations, with the tendency for the metal-rich populations to show a larger fractional contribution to the disturber than the metal-poor ones, though all of them have consistent vertical trends. The behaviors of $\langle V_R\rangle$ agree with the hypothesis that the in-plane substructures, mainly including the spirals and the bar, play roles in producing the structure of radial bulk velocities in the disk. Nonetheless, we cannot exclude the possibility that an external perturber, such as the passage of a satellite, causes the wobbly radial bulk motions.

Unlike the radial component of stellar bulk motion, the vertical component, which gets translated as a combination of breathing and bending mode motions, is stronger outside $R_0$ than inside. The breathing motion detected in the outer disk confirms the findings of \citet{Williams13} that there is a notable vertical contraction at $R>R_0$. The inner-disk rarefaction-like vertical motion, which was reported by some of the earlier works (e.g. \citealt{Williams13}; \citealt{Carrillo18}; \citealt{Carrillo19}), only appears at $|Z|>2\,\kpc$. The bending motion derived by outer-disk stars is aligned with the recent findings that an upward bending motion exists at $R>R_0$ (e.g. \citealt{Wang18}; \citealt{Carrillo19}; \citealt{Lopez-Corredoira20}; \citealt{Gaia Collaboration21}); by comparison, the downward bending motion inside $R_0$ is trivial. The directions for both the breathing and bending motions are mostly invariant with respect to $Z$, despite $Z$-dependent amplitudes.

The amplitude of the breathing motion depends on both $Z$ and $R$. The spiral density waves could drive alternate expanding and compressive motions along the Galactocentric radius near the plane (\citealt{Debattista14}; \citealt{Faure14}; \citealt{Ghosh20}), and contribute to an increasing amplitude of breathing motions with distance to the plane within the range $|Z|\lesssim1\,\kpc$ \citep{Ghosh20}. However, considering that the observed $|\vbreath|$ increases with the vertical height towards $|Z|\sim3\,\kpc$, it is unlikely that the breathing mode motion is totally attributed to spirals. As for the bending motion, \citet{Ghosh20} argued that the spirals are probably not responsible for the bending mode. \citet{Khoperskov19} proposed that the bending-mode oscillation could be induced by a buckling bar inhabited in the disk. The differences in vertical bulk velocities between different $\feh$ are trivial, indicating that different populations respond similarly to the perturbations. In this case, we infer that both the breathing and bending motions are produced or partly produced by long-lived external perturbations such as a satellite galaxy. It has been argued that the perturbed vertical motions could be generated by the Sagittarius dwarf-Milky Way interaction (\citealt{Gomez13}; \citealt{Widrow14}; \citealt{Antoja17}; \citealt{Laporte19}; \citealt{WangC19}).

The north-south asymmetries in velocity dispersions are detected in this work, though the differences between $Z>0$ stars and $Z<0$ stars are less pronounced than those for the mean velocities. It is interesting to find that the $\sigma_Z/\sigma_R$ in the south disk is larger than that in the north. The velocity dispersions are probes for the secular heating process in the disk, which are principally generated by spirals and giant molecular clouds (GMCs). The variation in $\sigma_Z/\sigma_R$ is a measure of the variation of the importance of the GMC as heating agents relative to spirals, since spirals heat mainly within the plane while GMCs heat both in-plane and vertically (\citealt{Jenkins92}; \citealt{Aumer16}; \citealt{Mackereth19}). The increasing $\sigma_Z/\sigma_R$ with increasing $|Z|$ indicates that the relative strength of spiral to GMC heating decreases with the distance to the plane within $|Z|\lesssim1.5\,\kpc$. Meanwhile, the larger $\sigma_Z/\sigma_R$ below the plane than above suggests either a weaker spiral heating agent or a larger number of GMCs at $Z<0$. Moreover, the interactions between a satellite galaxy and the disk could also heat the disk and cause north-south differences in velocity dispersions \citep{Mackereth19}.

The tilt of the velocity ellipsoid measured at $R<9\,\kpc$ is basically consistent with the classical relation case that the $\tilt$ is nearly antisymmetric about the plane, which was uncovered with various populations in the solar neighborhood (e.g. \citealt{Binney14}; \citealt{Budenbender15}; \citealt{Ding19}; \citealt{Everall19}). The $\tilt$ measured between $R=5$ and 9 kpc confirms the findings of \citet{Hagen19} that the orientation of the velocity ellipsoid is more aligned with spherical when we move closer to the GC. In order to compare the velocity ellipsoid orientation above and below the plane, we estimate the gradients of $\tilt$ with respect to $\arctan(Z/R)$ at $Z>0$ and $Z<0$ separately. We found a gradient of $0.803\pm0.026$, $0.784\pm0.019$ and $0.718\pm0.033$ at $5<R<7\,\kpc$, $7<R<8\,\kpc$ and $8<R<9\,\kpc$ respectively for the north disk, as well as $0.852\pm0.078$, $0.802\pm0.066$ and $0.321\pm0.075$ respectively for the south, suggesting that the flattening effect of the tilt along the Galactocentric radius is stronger in the south disk than in the north. The difference in the orientation between north and south disks is more pronounced in the Galactic anticenter. At $R>9\,\kpc$, when we move from $|Z|\lesssim1\,\kpc$ to higher north disk, the orientation of the velocity ellipsoid stays between a horizontal one and a spherical one. The reversed tilt observed at $Z<-1\,\kpc$, evidenced by populations with different $\feh$, indicates that the shapes of gravitational potential for the north and south parts of the outer disk are different when we move far away from the plane.

The observed vertical patterns of velocity ellipsoid help us to revisit the nature of the disk flaring. We find that the flaring of the disk begins near the solar position, which agrees with the findings of \citet{Gyuk99}. The position where the flaring starts to manifests itself is not settled. For instance, \citet{Lopez-Corredoira02} found a strong flare beginning well inside $R_0$ based on 2MASS red clump giants. To the contrary, \citet{LM14} and \citet{Yu21} found that the flare begins at $R\gtrsim10\,\kpc$ by using samples of SDSS-SEGUE F8V-G5V stars and LAMOST OB stars respectively. In recent studies, the flaring of a disk has been detected not only for our home galaxy but also for the external galaxies (\citealt{Borlaff16}; \citealt{Pinna19}; \citealt{Kasparova20}), which means that a flared disk could be a common structure in disk galaxies. The drivers of the formation of disk flaring is still unclear, one hypothesis is the minor mergers \citep{Bournaud09}.

The vertical structures of disk kinematics found in this work call for a deeper exploration of the origin of the long-lived north-south asymmetries in the disk. Besides, the shapes of the vertical patterns of velocity moments as functions of Galactocentric radius invite us to investigate the distribution of stellar kinematics in more dimensions in the future work. Additional measurements such as the data of chemical abundance provided by the ongoing medium-resolution survey of the LAMOST will give further insights into the Galactic structure and evolution.

\section{Summary}

In this work, we present an investigation into the vertical structure of stellar velocity moments at a Galactocentric radius of $R=5-15\,\kpc$ and 3 kpc above and below the Galactic plane in fine detail, based on K giant stars sampled from the LAMOST DR8. The distances of stars are estimated from the spectroscopically derived stellar parameters of LAMOST using the technique proposed by \citet{Carlin15}. The line-of-sight velocities are obtained from the LAMOST catalog, and the proper motions from {\it Gaia} EDR3. The metallicity of the sampled stars ranges from $\feh=-1$ to 0.5. The main results can be summarized as follows:

\begin{itemize}
  \item In the inner disk, stars both in the south and north disks are moving towards the Galactic anticenter, with mean velocities $\langle V_R\rangle$ increasing with the increasing distance to the plane. In the outer disk, there is an outward radial flow in the south disk and alternate outward and inward flows in the north; the $\langle V_R\rangle$ decreases generally with the increasing $R$, and the $\langle V_R\rangle$-$Z$ gradients become shallower for larger $R$. In the range of $8<R<12\,\kpc$, metal-rich populations have lower $\langle V_R\rangle$ than metal-poor ones.

  \item There is contraction-like breathing mode and upward bending mode motions in the outer disk. The amplitude of the breathing motion increases with the increasing $|Z|$ as a whole, up to around $10\,\kms$, and decreases towards the Galactic anticenter. The amplitude of the bending motion is typically within $5\,\kms$, and nearly vanishes when $|Z|$ approaches 3 kpc. By comparison, there is only weak or trivial rarefaction-like breathing mode and downward bending mode motions in the inner disk.

  \item The mean azimuthal velocity decreases with the increasing distance to the plane. When we move from the inner disk to the outer disk, the gradient $|d\langle V_\phi\rangle/dZ|$ decreases from $\sim27-28\,\kmskpc$ to $\sim5\,\kmskpc$. In the range of $5<R<12\,\kpc$, stars below the plane are rotating faster than stars above the plane. The $\langle V_\phi\rangle$-$\feh$ trend is positive at $5<R<7\,\kpc$, and turning negative at $13<R<15\,\kpc$.

  \item The velocity dispersions increase with the increasing distance to the plane and decrease with the increasing $R$. The $\sigma_Z/\sigma_R$-$Z$ trend is positive within $|Z|\lesssim1.5\,\kpc$ and hardly dependent on $R$. The $\sigma_Z/\sigma_R$ is on average larger at $Z<0$ than at $Z>0$. The tilt of the velocity ellipsoid is more aligned to a spherical orientation at smaller $R$, the gradient of $\tilt$ with respect to $\arctan(Z/R)$ increasing from $\sim0.52$ at $8<R<9\,\kpc$ to $\sim0.83$ at $5<R<7\,\kpc$. Within $9<R<15\,\kpc$, the velocity ellipsoid orientation lies somewhere between a horizontal one and a spherical one for $-1<Z<3\,\kpc$, while the vertical behavior of $\tilt$ reverses at $Z\lesssim-1\,\kpc$.

  \item The scaleheight of the disk increases from $\lesssim0.5\,\kpc$ at $R<8\,\kpc$ to $\gtrsim1.5\,\kpc$ at $R>12\,\kpc$, indicating a flaring feature on both sides of the disk. The south disk have larger $h_z$ than the north disk within $5<R<12\,\kpc$.
\end{itemize}

\acknowledgements
This study is supported by the National Natural Science Foundation of China under grants Nos. 11988101, 11873052, 11890694, 11773033, 11833004 and National Key R\&D Program of China No. 2019YFA0405500. The LAMOST FELLOWSHIP is supported by Special Funding for Advanced Users, budgeted and administrated by Center for Astronomical Mega-Science, Chinese Academy of Sciences (CAMS). Guoshoujing Telescope (the Large Sky Area Multi-Object Fiber Spectroscopic Telescope LAMOST) is a National Major Scientific Project built by the Chinese Academy of Sciences. Funding for the project has been provided by the National Development and Reform Commission. LAMOST is operated and managed by the National Astronomical Observatories, Chinese Academy of Sciences. This work has made use of data from the European Space Agency (ESA) mission {\it Gaia} (\url{https://www.cosmos.esa.int/gaia}), processed by the {\it Gaia} Data Processing and Analysis Consortium (DPAC, \url{https://www.cosmos.esa.int/web/gaia/dpac/consortium}). Funding for the DPAC has been provided by national institutions, in particular the institutions participating in the {\it Gaia} Multilateral Agreement.

\section*{Appendix}
The mean and dispersion of $V_R$ and $V_Z$ are estimated based on the two-dimensional Gaussian distribution. Let $\bm{V}=(V_R,\,V_Z)^T$ to be the observed velocity, and $\langle\bm{V}\rangle=(\langle V_R\rangle,\,\langle V_Z\rangle)^T$ to be the mean velocity. The probability density function (PDF) of $\bm{V}$ for one star is
\setlength{\mathindent}{0cm}
\begin{small}
\begin{equation}
f(\bm{V})=(2\pi)^{-1}\left|\bm{C_{V}}\right|^{-\frac{1}{2}}
\exp\left[-\frac{1}{2}(\bm{V}-\langle\bm{V}\rangle)^T\bm{C_{V}}^{-1}
(\bm{V}-\langle\bm{V}\rangle)\right].
\end{equation}\label{eqA1}
\end{small}

The covariance $\bm{C_{V}}$ is the sum of the dispersion tensor $\bm{D_{V}}$ and the covariance from the propagated observational uncertainties:
\begin{eqnarray}
\renewcommand\arraystretch{1.4}
\bm{D_{V}}&=&\left(
\begin{array}{cc}
\sigma^2_R &\sigma_{RZ} \\
\sigma_{RZ} & \sigma^2_Z
\end{array}
\right), \\
\bm{C_{V}}&=&\bm{D_{V}}
+\left(
\begin{array}{cc}
\delta V_R^2 &\rho \delta V_R  \delta V_Z \\
\rho \delta V_R \delta V_Z & \delta V_Z^2
\end{array}
\right),
\end{eqnarray}\label{eqA2}
where $\delta V_R$, $\delta V_Z$, and $\rho$ are the uncertainties in $V_R$, $V_Z$, and their correlation coefficient. For a group of $N$ stars, the log-likelihood distribution function for $\bm{V}$ is
\begin{equation}
L(\langle\bm{V}\rangle,\,\sigma_R,\,\sigma_Z,\,\sigma_{RZ})=\sum\limits_{i=1}^{N} \ln f_i(\bm{V}).
\end{equation}\label{eqA3}

The mean and dispersion of $\bm{V}$ can be estimated using the maximum-likelihood method.

\begin{figure}
\centering
\centerline{
\includegraphics[width=0.85\hsize]{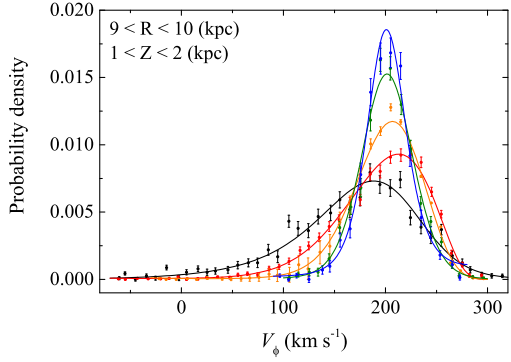}
}
\caption{The fits to the distribution of $V_\varphi$ for stars at $9<R<10\,\kpc$ and $1<Z<2\,\kpc$. The black, red, orange, olive, and blue dots/lines denote the metallicity range $-1<\feh<-0.6$, $-0.6<\feh<-0.4$, $-0.4<\feh<-0.2$, $-0.2<\feh<0$, $0<\feh<0.5$, respectively.}
\end{figure}\label{fig23}

Unlike $V_R$ and $V_Z$, the density function of $V_\phi$ is non-Gaussian. \citet{Binney14} proposed an analytic PDF for $V_\phi$, which has been used in \citet{Ding19} to estimate velocity moments for dwarf stars. The PDF of $V_\phi$, $f(V_\varphi)$, is written as
\begin{eqnarray}
f(V_\varphi)&=&\rm{constant} \times\exp \big(-\frac{(V_\varphi-b_0)^2}{2\sigma'^2_\varphi}\big),\nonumber\\
\sigma'_\varphi&=&b_1+b_2V_{\varphi100}+b_3V^2_{\varphi100}+b_4V^3_{\varphi100},
\end{eqnarray}\label{EqA4}
where $V_{\varphi100}\equiv V_\varphi/100\,\kms$. Since the normalizing constant is not available \citep{Binney14}, we estimate $b_0-b_4$ coupled with the normalizing constant by fitting the PDF to the observed $V_\varphi$ distribution. Figure 23 gives an example of the fits for populations with different metallicities. The mean and dispersion are derived by taking the first and second moments of the PDF.

\end{document}